\documentclass[aps,reprint,superscriptaddress]{revtex4-2}

\pdfoutput=1

\usepackage{color}
\usepackage{amsfonts}
\usepackage{amsmath}
\usepackage{graphicx}
\usepackage{textcomp}
\usepackage{hyperref}

\begin{document}


\title{Tunable vector beam decoder by inverse design for high-dimensional quantum key distribution with 3D polarized spatial modes}

\author{Eileen Otte}
\email{eileen.otte@stanford.edu}
\affiliation{Geballe Laboratory for Advanced Materials, Stanford University, Stanford, CA 94305, USA}

\author{Alexander D. White}
\affiliation{E. L. Ginzton Laboratory, Stanford University, Stanford, CA 94305, USA}

\author{Nicholas A. G\"{u}sken}
\affiliation{Geballe Laboratory for Advanced Materials, Stanford University, Stanford, CA 94305, USA}

\author{Jelena Vu\v{c}kovi\'c}
\affiliation{E. L. Ginzton Laboratory, Stanford University, Stanford, CA 94305, USA}

\author{Mark L. Brongersma}
\affiliation{Geballe Laboratory for Advanced Materials, Stanford University, Stanford, CA 94305, USA}

\date{\today}

\begin{abstract}
Spatial modes of light have become highly attractive to increase the dimension and, thereby, security and information capacity in quantum key distribution (QKD). So far, only transverse electric field components have been considered, while longitudinal polarization components have remained neglected. Here, we present an approach to include all three spatial dimensions of electric field oscillation in QKD by implementing our tunable, on-a-chip vector beam decoder (VBD). This inversely designed device pioneers the ``preparation'' and ``measurement'' of three-dimensionally polarized mutually unbiased basis states for high-dimensional (HD) QKD and paves the way for the integration of HD QKD with spatial modes in multifunctional on-a-chip photonics platforms. \\
\end{abstract}

\maketitle

\section{Introduction}

Fueled by Big Data and artificial intelligence (AI), there is an ever-increasing need for secure and high-performance data processing capabilities. This demand pushes the research on quantum computing as well as cryptography, which take advantage of the physics of light and quantum mechanics\,\cite{OBrien2009photonic}. While quantum computing could enable the desired fast and effective data processing, it would also enable faster breaking of standard encryption. Quantum-based encryption such as Quantum Key Distribution (QKD) tackles this issue: QKD facilitates two parties to securely share a cryptographic key, whereby the presence of an eavesdropper is revealed by implementing fundamental laws of quantum mechanics, namely, the no-cloning theorem\,\cite{shor2000simple}.

The best-known QKD scheme may be the BB84 protocol, introduced by Bennett and Brassard in 1984\,\cite{Bennett1984}. This prepare-and-measure protocol is based on using two bases of mutually unbiased basis (MUB) states -- by performing a single measurement, the sent quantum state cannot be fully identified if the encryption basis is unknown. Nowadays, researchers aim to further improve the performance of QKD, in particular, increasing the information capacity and transmission distances\,\cite{liao2017satellite, Boaron2018, Ding2017, Chen2022quantum}. In this context, high-dimensional (HD) QKD has attracted considerable attention since for an increasing dimension $d$ the information capacity per photon as well as the error threshold also increases\,\cite{bechmann2000quantum, ali2007large,  cerf2002security, Erhard2020advancesHD}. Providing access to larger dimensions by serving as MUB states, recently, in particular spatial modes of light have been of interest\,\cite{Otte2020Tutorial}. Orbital angular momentum (OAM)\,\cite{LSA7} as well as hybrid polarization-OAM modes have been implemented in HD QKD in free space, optical fibers, and underwater\,\cite{Ndagano2018, mirhosseini2015high, Hufnagel2020, Nape2018, Sit2017, cozzolino2019orbital, bouchard2018underwater}. Besides increasing $d$, using spatial modes comes along with the benefit of exploiting their inherent properties, e.g. the ability to self-reconstruct upon perturbation\,\cite{Nape2018}.

To efficiently implement spatial modes in QKD schemes, appropriate techniques for ``preparation'' and ``measurement'' of MUB states are required; in this context, future commercialization demands, in particular, stable, compact, and robust tools. So far, mainly bulky, complex techniques for encoding or decoding spatial modes have been realized, using spatial light modulators, digital mirror devices, combinations of q- and waveplates, or mode sorters\,\cite{berkhout2010efficient, Fickler2014quantum, mirhosseini2015high, Milione2015, Forbes2016, ndagano2017det, Nape2018, Fontaine2019laguerre, fickler2020full}. In contrast, integrated quantum photonics\,\cite{OBrien2009photonic, Wang2020integrated} holds the potential of miniaturizing and stabilizing encoding and decoding systems at low cost, scalability, and industrial compatibility. Additionally, on-a-chip prepare-and-measure devices -- QKD transmitters and receivers -- could be included in integrated platforms of multi-functionality, including optical processing of data.

Pioneered by the implementation of SiO$_2$-based optical interferometers for time-bin QKD in 2004\,\cite{Honjo2004differential}, different devices for time-bin as well as polarization-based QKD have been successfully introduced\,\cite{Sibson2017chip, Ma2016silicon, Sibson2017integrated, Bunandar2018metropolitan}. The usage of integrated photonics for HD QKD with spatial modes, however, is still in its infancy. To date, researchers proposed different integrated designs for the generation or detection of OAM modes and, more recently, polarization-structured vector modes\,\cite{Zhan2009cylindrical, Otte2020book} based on, e.g., ring-resonators, metasurfaces, grating or inverse-design structures\,\cite{Cai2012integrated, Su2012OAMdevice, Xie2018ultra, Zhou2019ultra, Chen2020vector, Huang2022leaky, Zheng2023versatile, White2022OAMEmitter}. However, most of these devices are static and thus limited in terms of their accessible modes and/or their implementation for HD QKD has not been considered -- partially because required MUB states cannot be prepared/measured.

Proposing an integrated prepare-and-measure tool for HD QKD based on scalar OAM and vector modes, we present an inversely designed, tunable device, we term a vector beam decoder (VBD). The VBD operates at the crucial interface of on-a-chip sender/receiver devices and free-space communication, thereby taking advantage of compact, multifunctional, scalable, integrated photonic platforms and satisfying the demand for long-distance information transmission. Notably, the VBD does not only enable the established HD QKD scheme which uses two-dimensionally (transversely) polarized spatial modes; it represents the first device, to the best of our knowledge, which facilitates the implementation of three-dimensionally (3D) polarized MUB states for HD QKD. The VBD prepares/ measures transverse as well as significant longitudinal electric field components, unlocking the still unexploited potential of 3D polarized spatial modes for QKD.

In the following, we first outline the concept of implementing spatial modes for HD QKD, in this case, 4D QKD (Sec.~\ref{subsec:HDQKD}). Subsequently, the tunable VBD is introduced in a back-projection scheme -- i.e. the VBD becomes a vector beam emitter (VBE) -- and we present accessible 2D polarized MUB states in the form of scalar OAM and vector modes (numerical results; Sec.~\ref{subsec:VBDasVBEandMUBstates}). Moving from 2D to 3D polarized MUB states, we analyze the compliance to the fundamental requirements for MUB states, the mode quality, as well as the security of the QKD scheme (Sec.~\ref{subsec:from2Dto3DMUB}). Finally, we discuss the presented results (Sec.~\ref{sec:DiscussionConclusion}).

\section{Results}

\subsection{Fundamentals on high-dimensional quantum key distribution with spatial modes} 
\label{subsec:HDQKD}

\begin{figure*}[tb]
 	\centering
 	\includegraphics[width=0.8\linewidth]{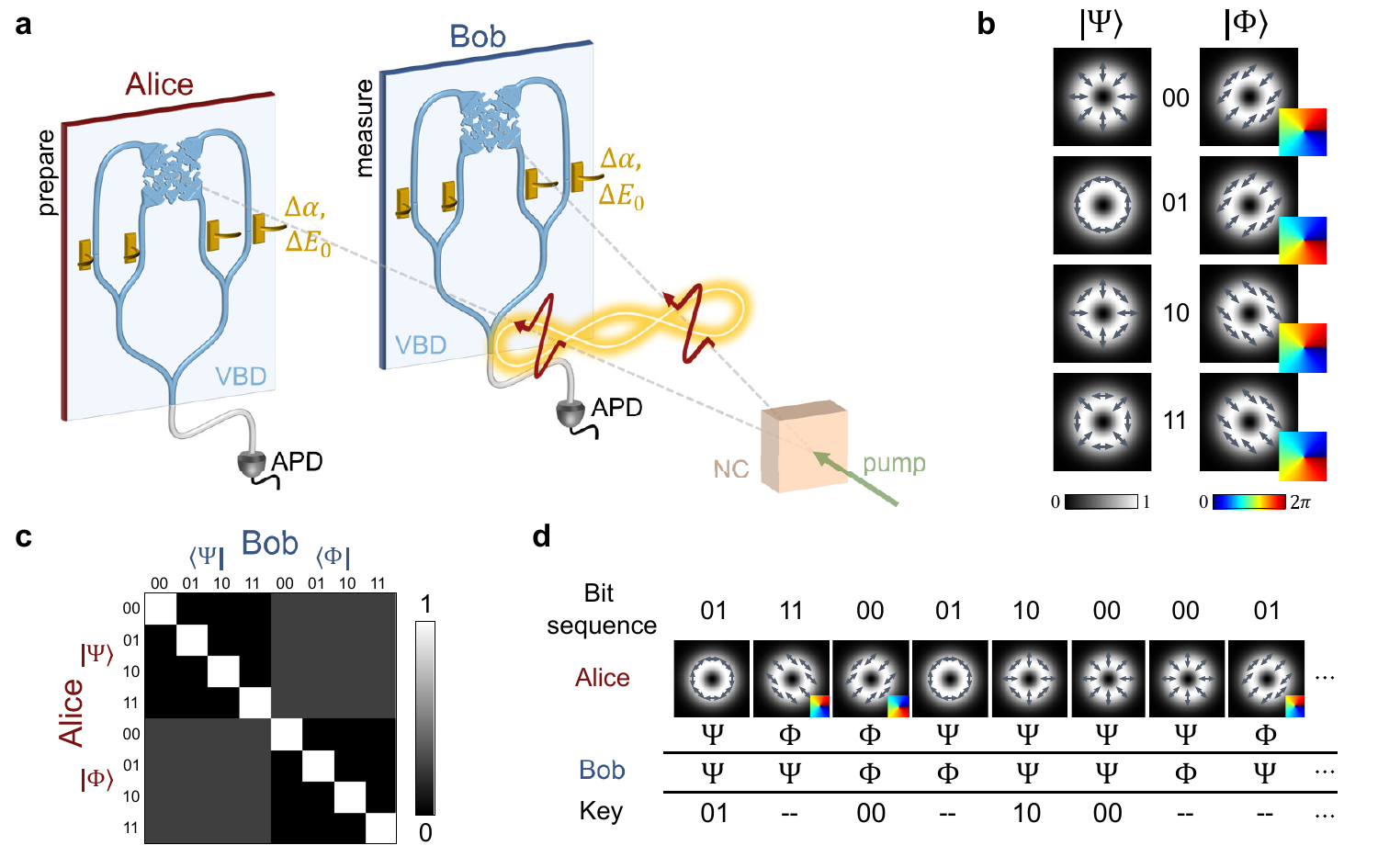}
 	\caption{The vector beam decoder (VBD) as an on-a-chip prepare-and-measure device for high-dimensional quantum key distribution (QKD). (a) Schematic representation of the experimental system (APD: avalanche photo diode; NC: nonlinear crystal; $\Delta \alpha$: adaptable relative phase shift; $\Delta E_0$: adaptable relative amplitude). (b) Mutually unbiased bases (MUB) $|\Psi\rangle$ and $|\Phi \rangle$ and their respective states, represented by spatial modes of light ($|\Psi\rangle$: vector modes; $|\Phi\rangle$: scalar modes). (c) Transfer matrix for communication between Alice and Bob. (d) Exemplary key generation process: Alice ``prepares'' her photons by her VBD in randomly selected bases to encode a bit sequence; Bob ``measures'' the entangled photons by his VBD in likewise randomly selected bases; bits sent in the same basis will be used as the key.}
 	\label{fig:Concept}
 \end{figure*}

Taking advantage of the fundamental laws of quantum mechanics, quantum key distribution (QKD) allows two parties, Alice and Bob, to securely exchange information while being able to detect the presence of an eavesdropper, Eve. In the QKD scheme of the BB84 protocol\,\cite{Bennett1984}, Alice and Bob unanimously agree on two orthonormal information bases $|\Psi\rangle$ and $|\Phi \rangle$; Alice and Bob randomly select one of these bases to ``prepare'' (Alice) or ``measure'' (Bob) the sent photon. This principle is depicted in Fig.~\ref{fig:Concept}a, where Alice and Bob interact each with one of two entangled photons, generated by pumping a nonlinear crystal (NC). While the first basis can be arbitrarily chosen with states $|\Psi_u\rangle,\: u = [1, d]\in \mathbb{N}$ and dimension $d$, the second basis must fulfill the condition:
\begin{equation}
    \vert\langle \Psi_u | \Phi_v\rangle \vert^2 = \frac{1}{d},
    \label{eq:MUBcondition}
\end{equation}
with $v = [1,\, d]\in \mathbb{N}$. This condition ensures that $|\Psi\rangle$ and $|\Phi \rangle$ are mutually unbiased bases (MUB). This means, when Alice randomly chose to sent her bits of information in basis $|\Psi\rangle$ ($|\Phi\rangle$), Bob detects the correct state with unity probability for measuring in the same basis $\langle \Psi|$ ($\langle \Phi|$), whereas he makes an ambiguous detection for measuring in the other basis. The same applies for Eve's measurements, which ultimately reveals her presence.

Originally, the communication (``prepare-and-measure'') protocol by Bennett and Brassard\,\cite{Bennett1984} is based on implementing 2D polarization states; more precisely, horizontal and vertical states ($|\Psi\rangle$) as well as diagonal and antidiagonal states ($|\Phi \rangle$) form the bases of this 2D QKD approach ($d=2$). To enhance the security of the key generation process\,\cite{cerf2002security}, we can increase the dimension $d$ by using the spatial degree of freedom (DoF) of photons. For this high-dimensional (HD) QKD, we implement spatial modes of light as MUB states.

An exemplary set of states for bases $|\Psi\rangle$ and $|\Phi \rangle$ is shown in Fig.~\ref{fig:Concept}b for $d=4$. In addition to the polarization DoF, spanning a 2D Hilbert space, here, we take advantage of the orbital angular momentum (OAM) DoF of a photon to create a 4D Hilbert space\,\cite{Otte2020Tutorial}. OAM modes of light are typically characterized by carrying a phase vortex structure, represented by the phase factor $\exp(\text{i} \ell \varphi)$ ($\varphi\in[0,\,2\pi]$: polar angle in a cylindrical coordinate system $(r,\varphi, z)$). Thereby, the topological charge $\ell$ describes the number of $2\pi$ cycles of the transverse phase vortex around the optical axis. The combination of the polarization DoF in the form of spin angular momentum (SAM), i.e. right- ($|R\rangle$) and left-circular ($|L\rangle$) polarization, and the OAM DoF with $\ell = \pm 1$ allows us to realize the first, so-called vector mode set\,\cite{ndagano2017det, Nape2018, Holleczek2011}:
\begin{equation}
\begin{split}
      | \Psi_{00}\rangle = \frac{1}{\sqrt{2}} E_0(r) [| R\rangle |\ell \rangle + | L\rangle |-\ell \rangle], \\
     | \Psi_{01}\rangle = \frac{1}{\sqrt{2}}  E_0(r) [|R\rangle |\ell \rangle - |L\rangle |-\ell \rangle],  \\
    | \Psi_{10}\rangle = \frac{1}{\sqrt{2}}  E_0(r) [|L\rangle |\ell \rangle + |R\rangle |-\ell \rangle], \\
    | \Psi_{11}\rangle = \frac{1}{\sqrt{2}}  E_0(r) [|L\rangle |\ell \rangle - |R\rangle |-\ell \rangle].
\end{split}
    \label{eq:Psi}
\end{equation}

As illustrated in Fig.~\ref{fig:Concept}b (left), these modes impart spatially varying linear polarization states in their transverse plane (gray arrows) and a ring shaped intensity structure, encoded in the amplitude $E_0(r)$. The set of MUB states is given by\,\cite{ndagano2017det, Nape2018, Holleczek2011}:
\begin{equation}
\begin{split}
     &|\Phi_{00}\rangle =  E_0(r) |D\rangle |-\ell \rangle , \\
     &|\Phi_{01}\rangle =  E_0(r) |D\rangle |\ell \rangle, \\
    &|\Phi_{10}\rangle = E_0(r) |A\rangle |-\ell \rangle , \\
    &|\Phi_{11}\rangle=  E_0(r) |A\rangle |\ell \rangle.
\end{split}
    \label{eq:Phi}
\end{equation}
As depicted in Fig.~\ref{fig:Concept}b (right), these scalar modes are of a homogeneous diagonal ($|D\rangle$) or antidiagonal ($|A\rangle$) linear polarization (gray arrows) and carry a phase vortex of positive or negative topological charge $|\ell |  =1$ (insets).

Per basis, each state corresponds to another bit of information, namely, $00$, $01$, $10$, or $11$, which Alice can prepare to sent to Bob. Fig.~\ref{fig:Concept}c presents the ideal transfer matrix for the communication between Alice and Bob for implementing the spatial modes in Fig.~\ref{fig:Concept}b. It shows the probability of Bob identifying the correct state sent by Alice. If the preparation and measurement bases match, Bob can unambiguously identify the sent state; for a mismatch, Bob makes an ambiguous detection, measuring the correct state only with a probability $1/d = 25\%$.

The key generation process for this 4D QKD approach is sketched in Fig.~\ref{fig:Concept}d. Alice randomly selects a basis to sent a bit sequence of length $N$ to Bob; each information bit corresponds to another scalar or vector state. Bob chooses his measuring basis at random, too. At the end of the transmission, Alice and Bob compare their bases selection publicly via a classical communication channel and discard bits measured in the instances of bases mismatch. On average, Alice and Bob should be left with $N/2$ information bits. However, if an eavesdropper is present, these bit would contain errors induced by Eve's measurements. By publicly comparing a subset of the sifted $N/2$ bits, Alice and Bob can reveal Eve's presence, estimate the induced error, and decide whether to proceed with the rest of the sifted key or discard it.

\subsection{The tunable vector beam decoder as a prepare-and-measure device}
\label{subsec:VBDasVBEandMUBstates}

For HD QKD with spatial modes, Alice and Bob need to apply a spatially resolved preparation/ measurement technique, as indicated in Fig.~\ref{fig:Concept}a. Working at the topical interface of free-space and integrated optical communication, we inversely designed a compact, integrated vector beam decoder (VBD) as ``prepare-and-measure'' tool. This tool can decode, hence, prepare and measure the spatial modes presented in Fig.~\ref{fig:Concept}b. To outline the working principle of the VBD, let us consider the QKD system (Fig.~\ref{fig:Concept}a) in the case of classical back-projection\,\cite{klyshko1988simple}. For classical back-projection, the NC is replaced by a mirror and Alice's detector (APD: avalanche photo diode) by a classical light source. This source sends photons backwards through the VBD, thereby transforming it into a tunable vector beam emitter (VBE; orange: phase-and-amplitude modulators). The light field, which is emitted orthogonally to the on-a-chip VBE, propagates to Bob, after being reflected by the mirror in the NC plane. Bob measure the light field by his VBD, detecting its output by his APD.

\begin{figure*}[htb]
 	\centering
 	\includegraphics[width=0.8\linewidth]{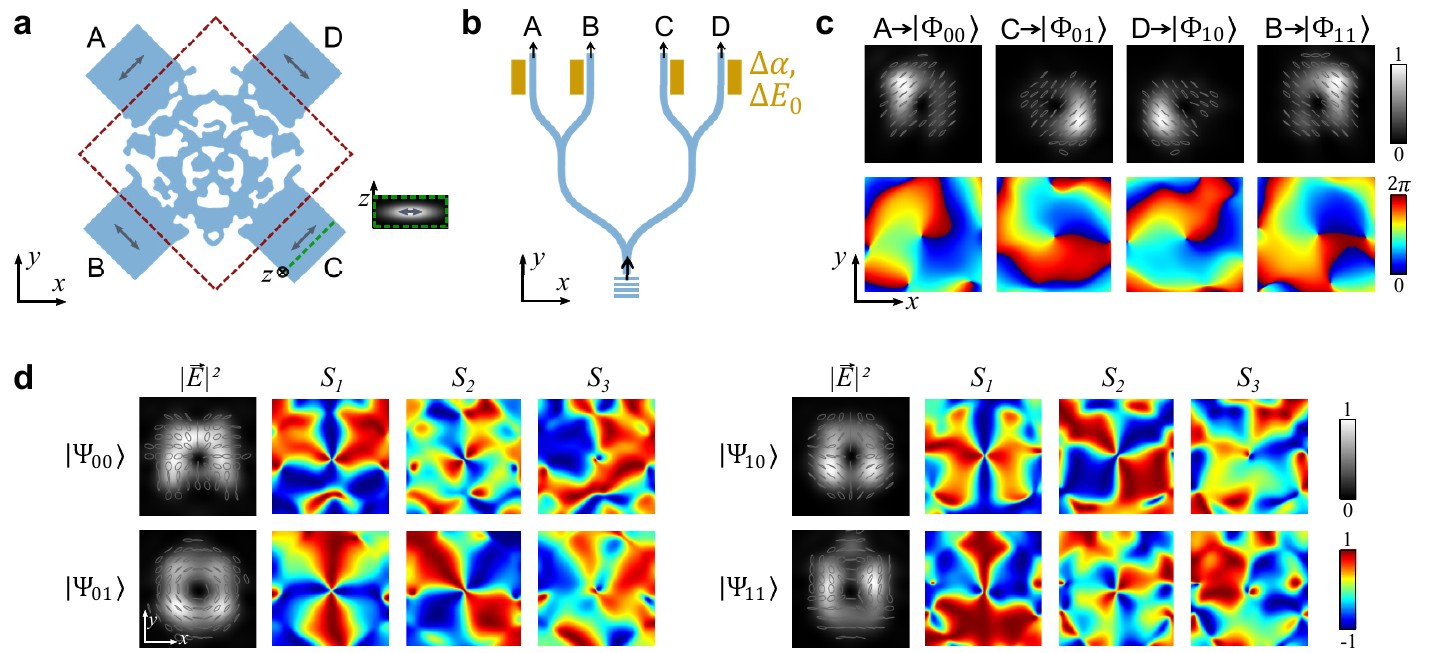}
 	\caption{Generating MUB states by the VBD used as vector beam emitter (VBE) (classical back-projection approach). (a) VBE with inverse design area (red dashed square) and its input waveguides A, B, C, and D. Linear polarization per waveguide is indicated by a gray arrow; waveguide mode in green dashed plane (input C) is illustrated as an inset (green dashed box); the output of the VBE propagates in the $z$-direction. (b) Sketch of tunable input waveguides with a coupling grating (bottom) as well as phase-and-amplitude modulators (orange; $\Delta \alpha_\text{A,B,C,D}\in [0, \, 2\pi]$; $\Delta E_\text{0; A,B,C,D}\in [0,\,1]$). (c) MUB states of $|\Phi\rangle$ generated by input A, C, D, and B; top: (transverse) polarization ellipses on intensity $|\vec{E}|^2 = |\Phi_{ij}|^2\in [0,\,1]$, bottom: phase distribution arg$(\Phi_{ij})\in [0,\, 2\pi], \: i,j = \lbrace 0,\, 1\rbrace$. (d) MUB states of $|\Psi\rangle$ realized by a combination of different input channels A-D, phase shifts $\Delta \alpha_\text{A-D}$, and amplitude adaptions by $\Delta E_\text{0; A-D}$. Per state, left to right: (transverse) polarization ellipses on intensity $|\vec{E}|^2 = |\Psi_{ij}|^2\in [0,\,1],\: i,j = \lbrace 0,\, 1\rbrace$ and respective normalized Stokes parameters $S_{1,2,3}\in [-1, 1]$.}
 	\label{fig:DeviceMUBs}
 \end{figure*}

The VBE is sketched in more detail in Fig.~\ref{fig:DeviceMUBs}a, b. The tunable VBE is comprised of an inversely designed coupling region (Fig.~\ref{fig:DeviceMUBs}a, red dashed box) with four input waveguides A, B, C, and D, which combine into a single single-mode waveguide (Fig.~\ref{fig:DeviceMUBs}b, simplified sketch). Light is coupled into the VBE by a coupling grating. The inversely designed region as well as waveguides and the coupling grating are patterned into an air-clad $220\,\mbox{nm}$ silicon on insulator (details in Methods). When joining into the inverse-design area, the waveguides of rectangular cross section support a single mode of elliptical shape (green dashed line/box) and, dependent on waveguide orientation, diagonal or antidiagonal linear polarization (gray arrows). Each of the waveguide modes can be shifted in phase by $\Delta \alpha_\text{A-D}$ by electrically controlled gold (Au) heaters (orange, Fig.~\ref{fig:DeviceMUBs}b). By using Michelson interferometers (not sketched here; see Methods), we are also able to control the relative amplitude $\Delta E_\text{0;A-D}$ of the input waveguides.

The VBE is designed in such a way that each of the input waveguides A-D generates an output light field $\vec{E}_\text{A-D} = [E_x, E_y, E_z]^T$ with its transverse components $\vec{E}_\perp = [E_x, E_y]^T$ representing one of the four scalar MUB states $|\Phi_{ij}\rangle,\, i,\,j = \{1,\,0\}$. These output modes are presented in Fig.~\ref{fig:DeviceMUBs}c (simulation), with the top row showing the transverse polarization states (gray) upon the total intensity $|\vec{E}|^2$ (normalized; grayscale) and the bottom row depicting the respective transverse phase structure arg$(\Phi_{ij})$ of (anti)diagonally polarized components. 

The MUB states $|\Psi_{ij}\rangle$ can be realized by using a combination of different input waveguides (A-D), phase shifts $\Delta \alpha_\text{A-D}$, and amplitude adaptions $\Delta E_\text{0;A-D}$. Simplified, this can be explained as follows: $|R\rangle \propto |D\rangle + \text{i} |A\rangle$ and $|L\rangle \propto |D\rangle - \text{i} |A\rangle$. Hence, $|R\rangle$ and $|L\rangle$ polarized parts of each vector state $|\Psi_{ij}\rangle$ (Eq.~(\ref{eq:Psi})) can be formed by combining two phase-shifted ($\Delta \alpha$) orthogonally polarized $|\Phi_{ij}\rangle$ states of equal topological charge $\ell$, i.e. two different input waveguides are used. The tunable phase shifts $\Delta\alpha_\text{A-D}$ further enable us to combine newly formed $|R\rangle$ and $|L\rangle$ polarized parts according to Eq.~(\ref{eq:Psi}). Amplitude adaptions $\Delta E_\text{0;A-D}$ can be used for normalization and function as on-off switch per waveguide. Following this principle, the vector modes $|\Psi_{ij}\rangle$ in Fig.~\ref{fig:DeviceMUBs}d (simulation) can be formed. Per state, from left to right, we present the transverse polarization (gray) upon the total intensity $|\vec{E}|^2$ (normalized), and normalized Stokes parameters $S_{1,2,3} \in [-1,\,1]$. The Stokes parameters $S_{1,2,3}$ represent the amount of horizontal/vertical, diagonal/antidiagonal, or right-/left-circular polarized components of a transverse electric field, respectively. For vector modes of basis $|\Psi\rangle$, $S_{1,2,3}$ highlight the spatial variation in transverse polarization.

Clearly, the VBE supports the desired modes of light, thus seem to fulfill the requirements for the outlined HD QKD approach. To study the quality of these modes, thereby analyzing the security of the respective key generation process, we will subsequently take a look at the transfer matrix and selected security values. Also, the orthonormality and fundamental requirement of MUB will be analyzed for the 3D electric field of the VBE/VBD.

\subsection{From 2D to 3D polarized MUB states: mode quality, security analysis, and 3D states}
\label{subsec:from2Dto3DMUB}
 
To quantify the security of the QKD approach based on implementing the VBE (i.e. VBD in the quantum system), we investigate the normalized transfer matrix $T$ (see Methods). To simultaneously determine the quality of realized MUBs $|\Psi_{ij}\rangle$ and $|\Phi_{ij}\rangle$ (simulations), we first assume Alice sends her bits using the inversely designed VBE while Bob uses idealized modes $\langle \Psi_{ij}|$ and $\langle \Phi_{ij}|$ following the complex conjugated Eqs.~(\ref{eq:Psi}) and (\ref{eq:Phi}), respectively. Thereby, $E_0(r)|\pm\ell\rangle$ is represented by a helical Laguerre-Gaussian (LG) mode\,\cite{Siegman1986, Boyd1961} of topological charge $\ell = \pm 1$ and radial index $p=1$; its beam waist is matched to Alice's mode size. Note that, in this case, Bob's decoding basis is 2D polarized, hence, only $\vec{E}_\perp = [E_x,E_y]^T$ of the VBE is considered for QKD and analyzed for its mode quality.

\begin{figure}[bt]
 	\centering
 	\includegraphics[width=\linewidth]{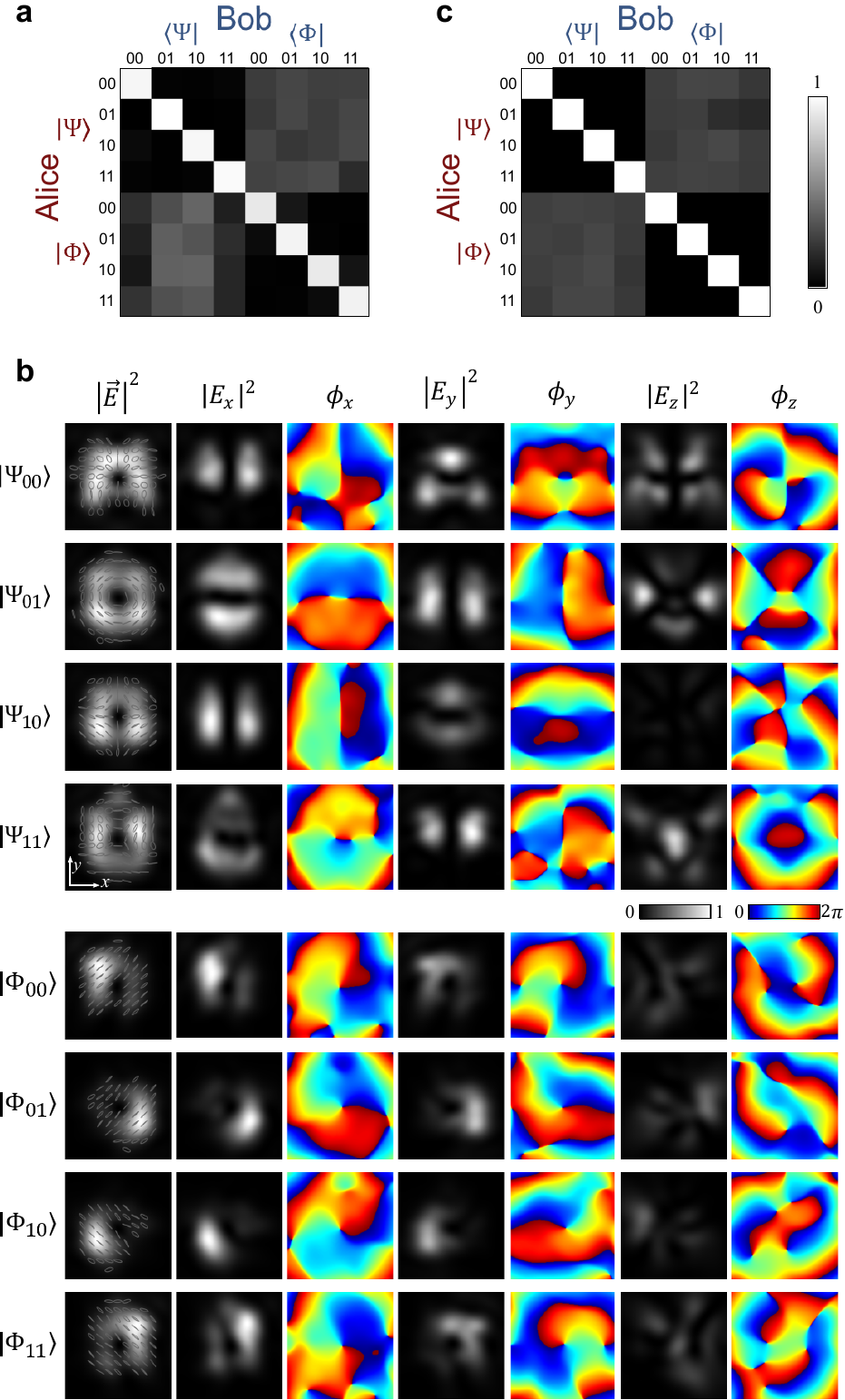}
 	\caption{Security analysis for the key generation by Alice and Bob (classical back-projection) based on 2D and 3D polarized MUB states. Transfer matrices for the cases that (a) Bob's measurement is idealized by using complex conjugated Eqs.~(\ref{eq:Psi}) and (\ref{eq:Phi}) as 2D decoding bases ($E_z = 0$ for $\langle \Psi|$ and $\langle \Phi|$), or (c) his measurement is performed by the VBD, taking the whole 3D electric field into account. (b) The intensity $|E_{n}|^2$ (normalized to the maximum of $|E_{x,y,z}|^2$) and phase $\phi_n, \: n = \lbrace x,y,z \rbrace$ per polarization component of $|\Psi_{ij}\rangle$ or $|\Phi_{ij}\rangle$.}
 	\label{fig:TransferMatrices3DFields}
 \end{figure}

In Fig.~\ref{fig:TransferMatrices3DFields}a we depict the normalized transfer matrix, considering idealized MUB states for Bob's measurement. In Table~\ref{tab:SecurityAnalysis}, we present the respective security values and compare the quantum error rate $Q$, detection fidelity $F$, the mutual information $I_{AB}$ between Alice and Bob, Eve's cloning fidelity $F_E$, her mutual information $I_{AE}$ with Alice, and the information capacity per photon $R_\Delta$ in bits/photon to the ideal values in case of 4D QKD (details on security values are described in the Methods).

The top left and bottom right quadrant of the transfer matrix (Fig.~\ref{fig:TransferMatrices3DFields}a) correspond to $|\langle \Psi_u | \Psi_v \rangle|^2$ and $|\langle \Phi_u | \Phi_v \rangle|^2$, respectively, hence, allow imminent conclusions about quality of the VBE's MUB states and their orthonormality per basis. For $u=v$ (diagonal matrix elements), matrix elements reveal the qualitatively good match between Bob's idealized 2D polarized decoding basis states and Alice's states, realized by the VBE. For $u\neq v$, values are close to zero, demonstrating the desired unambiguity for measuring Alice's states in the correct basis. This observation is quantified in the quantum error rate $Q= 0.057$ and measurement fidelity $F = 0.943$, being close to the ideal values of $0$ and $1$, respectively. For measurements in the case of bases mismatch, i.e. $|\langle \Psi_u | \Phi_v \rangle|^2$ or $|\langle \Phi_u | \Psi_v \rangle|^2$ (top right/ bottom left quadrant), the transfer matrix reveals small perturbations in the probability values. Ideally, the value is $1/d = 0.25$ (cf. Fig.~\ref{fig:Concept}c; Eq.~(\ref{eq:MUBcondition})). Nevertheless, the slightly perturbed transfer matrix allow us to calculate very convincing security values:

Dependent on $Q$, the amount of information that Alice and Bob share in this QKD system, can be calculated, i.e. $I_{AB} = 1.594$ bits per photon (ideal value is $2$). Eve's cloning fidelity, i.e. her ability to copy a sent quantum state, is quantified by $F_E = 0.479$, reflecting the perturbations observed in the transfer matrix. Impressively, even though this value deviates from the ideal one ($0.25$), the determined cloning fidelity is still lower than the value in the standard 2D QKD case, i.e. $F_E^\text{2D} = 0.5$. The associated mututal information shared between Alice, using the VBE, and Eve is calculated as $0.239$ (ideal value is $0$). Finally, the secret key rate $R_\Delta$ of value $1.189$ is above the Shannon limit of one bit per photon achievable with qubit states and, thus, demonstrates the benefit of implementing spatial modes by the VBE for realizing HD QKD.

\begin{table}[hb]
\centering
	\caption{Security analysis of the QKD transfer matrices in Fig.~\ref{fig:TransferMatrices3DFields}. The respective quantum error rate $Q$, detection fidelity $F$, the mutual information $I_{AB}$ between Alice and Bob, Eve's cloning fidelity $F_E$, her mutual information $I_{AE}$ with Alice, and the information capacity per photon $R_\Delta$ in bits/photon is compared to the ideal values in case of 4D QKD.} \vspace{0.4em}
		\begin{tabular}{c | c c c c c c }
			 & $Q$ & $F$ & $I_{AB}$ & $F_E$ & $I_{AE}$ & $R_\Delta$ \\ \hline \hline 
			Ideal & $0$ & $1$ & $2$ & $0.25$ & $0$  & $2$\\
			Fig.~\ref{fig:TransferMatrices3DFields}a, 2D MUB \ & \ $0.057$ & $0.943$ & $1.594$ & $0.479$ & $0.239$ & $1.189$\\
                Fig.~\ref{fig:TransferMatrices3DFields}c, 3D MUB \ & \ $0.004$ & $0.996$ & $1.957$ & $0.306$ & $0.018$ & $1.914$\\
		\end{tabular}
	 \label{tab:SecurityAnalysis}
\end{table}

Crucially, these security values can be improved even further, becoming close to ideal, when considering the whole 3D polarized electric field of the VBE. As presented in Fig.~\ref{fig:TransferMatrices3DFields}b (simulations), the modes emitted by the VBE actually carry a significantly contributing longitudinally polarized component $E_z$. For each MUB state of bases $|\Psi\rangle$ and $|\Phi \rangle$, from left to right, we depict the transverse polarization $\vec{E}_\perp = [E_x,E_y]^T$ (gray) upon the total field intensity $|\vec{E}|^2\in[0,\,1]$ with $\vec{E} = [E_x, E_y, E_z]^T$, and, per polarization component, its intensity contribution $|E_n|^2\in[0,\,1]$ as well as its phase distribution $\phi_n =\text{arg}(E_n)\in[0,\,2\pi]$, $n=\{x,\,y,\,z\}$. The significance of $E_z$ is particularly visible for vector modes of basis $|\Psi\rangle$. Note that, at the single-photon level, such as in the case of the intended quantum experiment, each photon obviously only shows electric field oscillation perpendicular to its propagation direction. Photons of 3D polarization can be understood as photons propagating under an angle with respect to the assigned optical axis (here: $z$-axis) of the considered coordinate system. For transmitting photons of these modes in free space, a lens would perform the task of bridging between non-paraxial (VBD/VBE) and paraxial regime (free space) -- without the loss of information. Following the typical implementation of spatial modes in QKD, the prepare-and-measure plane ($=$ VBD plane) would be imaged onto the nonlinear crystal plane (NC in Fig.~\ref{fig:Concept}a; mirror in classical back-projection).

Considering these additional polarization components in the QKD approach, we determine the normalized transfer matrix for Alice and Bob both using the VBE/VBD; the matrix is presented in Fig.~\ref{fig:TransferMatrices3DFields}c. We use this matrix to verify that the 3D polarized modes fulfill the fundamental conditions for MUB states, and determine respective security values. Obviously, the matrix shows less perturbations than in Fig.~\ref{fig:TransferMatrices3DFields}a, with particular improvement in the matrix quadrants of basis mismatch (top right, bottom left). Importantly, from these matrix values (mean value $0.244\approx 1/d$), we can conclude that the condition in Eq.~(\ref{eq:MUBcondition}) is, on average, fulfilled for our 3D polarized MUB states. Table~\ref{tab:SecurityAnalysis} shows the calculated security values for measuring with the VBD, thus, in case of implementing the 3D polarized MUB states in the QKD approach. Overall, the security values only insignificantly deviate from the ideal values, proofing the quality and the feasibility of the inversely designed VBE/VBD for HD QKD. Further, these results present the first implementation of 3D polarized MUB states for HD QKD and, simultaneously, the huge potential of this implementation.

\section{Discussion \& Conclusion}
\label{sec:DiscussionConclusion}

The inversely designed, tunable VBE enables the on-demand generation of scalar and vector modes of light. Implemented as VBD, this integrated prepare-and-measure device enables, for the first time to the best of our knowledge, the implementation of 3D polarized MUB states for 4D QKD of very high security (cf. Tab.~\ref{tab:SecurityAnalysis}). The VBD was designed such that, for transverse polarized components, it matches the set of known vector and scalar modes, previously implemented for HD QKD\,\cite{ndagano2017det, Nape2018} (Eqs.~(\ref{eq:Phi}), (\ref{eq:Psi})). By choosing to create an integrated, on-a-chip device that is oriented perpendicular to the optical axis of incoming photons ($z$-axis $\perp (x,\,z)$-plane), longitudinal electric field components take a significant role. This choice allows us to create the new set of 3D polarized MUB states.

As a compact, on-a-chip device, the VBD has the potential to be implemented in future quantum photonic integrated circuits\,\cite{OBrien2009photonic, Wang2020integrated}. This topical area of research is motivated by the scalability, stability, and multifunctionality of on-a-chip devices. In such a circuit, the VBD can not only serve as a prepare-and-measure device for HD QKD, but could also be used for optical quantum communication with spatial-mode based multiplexing\,\cite{Milione2015}. In both cases, the VBD works at the crucial intersection of free-space and on-chip communication. Although spatial modes have proven beneficial for data multiplexing and HD QKD, especially in free space, their implementation within integrated circuits or optical fibers remains challenging. By combining the benefits of both, integrated circuits and spatial modes of light, the designed VBD bridges between free-space transmission and integrated processing of optical data at the classical and quantum level.

Standard optical decoding systems for spatial modes, for QKD as well as for communication, only consider the transverse electric field components of light as a valid approximation in the paraxial regime. However, the third, longitudinal polarization component holds tremendous untapped potential: they can enable completely new sets of spatial modes for data encoding and, thereby, broaden the spectrum of applicable modes. Also, similar to OAM, $z$-components and their properties might be considered as an additional DoF of photons, with respect to which quantum superposition states can be formed. To facilitate these exciting future direction of implementing 3D fields, we require encoding as well as decoding devices for classical as well as quantum optics, which can handle 3D polarized light -- the VBD/VBE represents the first tool of this kind.

\begin{acknowledgments}
E.O. acknowledges support by the DAAD PRIME Program as well as the Geballe Laboratory for Advanced Material Postdoctoral Fellowship (GLAM Fellowship), Stanford University. A.D.W. acknowledges the Herb and Jane Dwight Stanford Graduate Fellowship (SGF) and the NTT Research Fellowship for support. N.A.G. thanks the German National Academy of Sciences Leopoldina for their support via the Leopoldina Postdoctoral Fellowship (LPDS2020-12). M.L.B. acknowledges support from the United States Air Force Office of Scientific Research under a MURI grant (Award: \#FA9550-21-1-0312).
\end{acknowledgments}

\onecolumngrid

\section*{Methods} 
\label{sec:Methods}

\subsection{Representation of MUB states by Laguerre-Gaussian modes}

The 2D polarized MUB states (scalar and vector modes) in Sec.~\ref{subsec:HDQKD} and Eqs.~(\ref{eq:Psi}), (\ref{eq:Phi}) can be described as a superposition of helical Laguerre-Gaussian modes $\text{LG}_{p,\ell}$. These modes represent exact analytical solutions to the paraxial wave equation in polar coordinates $(r, \varphi, z)$ with its complex amplitude\,\cite{Siegman1986, Boyd1961}:
\begin{align} 
    & \text{LG}_{p,\ell}(r,\,\varphi ,\,z) = A_{p,\ell}(r,z) \cdot \text{e}^{\text{i} \frac{k r^2}{2R(z)}}\cdot \text{e}^{\text{i}\phi_{p,l}^{G}(z)}\cdot \text{e}^{\text{i}\ell \varphi}, \quad\text{and}\label{eq:LGbeam}\\
    &A_{p,\ell}(r,z) = \sqrt{\frac{2p!}{\pi(|\ell|+p)!}}\cdot \frac{1}{w(z)}\cdot \text{e}^{-\frac{r R{^2}}{w^2(z)}} \cdot \left(\frac{r\sqrt{2}}{w(z)}\right)^{|\ell|}\cdot \text{L}_p^{|\ell|}\left(\frac{2r^2}{w^2(z)}\right),\\
    &\phi_{p,\ell}^G(z) = (2p+|\ell|+1) \, \phi_{0,0}^G(z)
\end{align}
($p\in \mathbb{N}_0$: radial mode number, $k$: wave number, $R(z)$: wave front curvature, $w(z)$: beam radius, $w_0=w(0)$: beam waist). $L_p^\ell(\cdot)$ represents the eponymous Laguerre polynomial\,\cite{Siegman1986, Boyd1961}, and $\phi_{p,\ell}^G$ the Gouy phase shift of LG modes ($\phi_{0,0}^G$: Gouy phase of fundamental Gaussian beam). In the context of this work, we consider $E_0(r) |\pm \ell\rangle = \text{LG}_{p=0,\pm\ell}(z=0)$. For this assumption, vector modes Eq.~(\ref{eq:Psi}) are also known as cylindrical vector beams\,\cite{Zhan2009cylindrical}.

\subsection{Inverse design of the VBD}

We use adjoint optimized photonics inverse design to create the VBE, which can be implemented as VBD. For the design, we aim for the MUB states of Eqs.~(\ref{eq:Psi}) and (\ref{eq:Phi}) to be formed in the transverse electric field components $\vec{E}_\perp = [E_x,\,E_y]^T$ of the VBE emission. For this purpose, we assume the MUB states to be composed of LG modes $\text{LG}_{p=0,\pm\ell}$ (see above). We optimize an emitter structure that takes a waveguide mode (lowest-order TE mode) as input and creates a beam out of the emitter plane ($(x,y)$-plane) which maximally overlaps with the desired electric field in free space. For realizing all MUB states with the same emitter structure, we assume four input waveguides with each of them resulting in one of the four $|\Phi\rangle $ states as output. By implementing phase and amplitude modulation per waveguide, these four output modes also allow us to create all MUB states of basis $|\Psi\rangle$.

The emitter structure is $220\,\mbox{nm}$ thick, is $3 \times 3 \,\mbox{\textmu m}$ in transverse size (Fig.~\ref{fig:DeviceMUBs}a, red dashed box), and has $1.5\,\mbox{\textmu m}$ wide input waveguides. By an adjoint optimization approach, we efficiently calculate the optimization gradients at every point with only two simulations\,\cite{Goos2021, Su2020nanophotonic}. Thereby, we maximize the overlap of the output field and the wanted MUB mode, i.e.
\begin{equation}
    \max_{\varepsilon} |\vec{c}^{\,\dag} \, \vec{e}(\varepsilon)|^2.
\end{equation}
Here, $\vec{c}$ is the vectorized electric field of the desired MUB state and $\vec{e}(\varepsilon)$ represents the vectorized electric field due to the designed-permitted distribution. The starting condition was a circle of radius $1.5\,\mbox{\textmu m}$ to best support cylindrical symmetric LG modes. The resulting emitter structure is shown in Fig.~\ref{fig:DeviceMUBs}a.

\subsection{Fabrication design for a tunable VBE}

To enable the tunable performance of the VBE/VBD, we incorporate phase and amplitude modulation per waveguide A, B, C, and D. For this purpose, waveguides form Michelson interferometers on the device chip; per waveguide, i.e. interferometer ``arm'', the relative phase can be dynamically adapted between $0$ and $2\pi$ by gold (Au) heaters, placed next to each waveguide. The heaters allow a local refractive index change in the respective waveguide, resulting in a phase retardation for the transmitted light.
For functioning as the VBE, light can be coupled into the device by a coupling grating at a joint input waveguide (cf. sketch in Fig.~\ref{fig:DeviceMUBs}b). The grating is designed according to ref.~\cite{Sapra2019inverse}.

\subsection{Calculation of security values}

The security analysis\,\cite{cerf2002security, Otte2020Tutorial} is performed from the transfer matrix $T$, with its four quadrants being described by:
\begin{equation}
T = \begin{pmatrix}
|\langle{\Psi^{\text{Bob}}_u}|\hat{U}|{\Psi^{\text{Alice}}_v}\rangle|^2	& |\langle{\Psi^{\text{Bob}}_u}|\hat{U}|{\Phi^{\text{Alice}}_v}\rangle|^2 \\
|\langle{\Phi^{\text{Bob}}_u}|\hat{U}|{\Psi^{\text{Alice}}_v}\rangle|^2& |\langle{\Phi^{\text{Bob}}_u}|\hat{U}|{\Phi^{\text{Alice}}_v}\rangle|^2
\end{pmatrix},\quad u,\,v = \{00,\, 01,\, 10,\, 11\}.
\label{eq:TranferMatrix}
\end{equation}
Here, the operator $\hat{U}$ represents the communication channel, through which Alice sents her states to Bob. In the frame of this work, we assume an idealized channel with $\hat{U}=1$. For determined matrices (Fig.~\ref{fig:TransferMatrices3DFields}a,b), each row $m=\{1,..,8\}$ is normalized by the sum of $|\langle{\Psi^{\text{Bob}}_u}|{\Psi^{\text{Alice}}_m}\rangle|^2$ or $|\langle{\Phi^{\text{Bob}}_u}|{\Phi^{\text{Alice}}_m}\rangle|^2$ values, respectively.

The transfer matrix components allow us to determine the quantum error rate $Q$ as the mean of:
\begin{eqnarray}
Q_{\Psi} &=& 1 - \frac{1}{2}\sum_{i=1}^{2} |\langle{\Psi^{\text{Bob}}_i}|{\Psi^{\text{Alice}}_i}\rangle|^2, \label{eq:error_rate1}\\
Q_{\Phi} &=& 1 - \frac{1}{2}\sum_{i=1}^{2} |\langle {\Phi^{\text{Bob}}_i}{\Phi^{\text{Alice}}_i}\rangle|^2, \label{eq:error_rate2}
\end{eqnarray}
with the second term representing the measurement fidelity $F$. Based on $Q$ ($F$) and the dimension $d$ (here, $d=4$), we can calculate further security values\,\cite{cerf2002security}, such as the mutual information between Alice and Bob, i.e.:
\begin{equation}
 I_{AB}(d,Q) = \log_2(d) + (1-Q)\log_2(1-Q)+ (Q)\log_2\left(\frac{Q}{d-1}\right).
\label{eq:mutalinfo}
\end{equation}
Noticeably, more information can be packed into every single photon with increasing dimension $d$. The choice of dimension as well as the quantum error rate $Q$ of the system also affect the eavesdropper cloning fidelity:
 \begin{equation}
 F_E(d,Q) =\frac{1}{d}\left(1 + (d-2)Q + 2\sqrt{(d-1)Q(1-Q)}\right).
  \label{eq:evefidelity}
 \end{equation}
The respective mutual information shared between Alice and Eve is given by:
 \begin{equation}
I_{AE}(d,Q)  = \log_2(d)+(F_E - Q)\log_2\left(\frac{F_E-Q}{1-Q}\right) + (1-F_E)\log_2\left(\frac{1-F_E}{(d-1)(1-Q)}\right).
 \label{eq:AliceEveInfo}
\end{equation}
For a two bases protocol ($|\Psi\rangle$ and $|\Phi\rangle $) in higher dimensions, the secret key rate is described by\,\cite{Sheridan2010}:
\begin{equation}
R_\Delta(d,Q) = \log_2(d)+2(1-Q)\log_2(1-Q)+2Q\log_2\left(\frac{Q}{d-1}\right).
\label{eq:secretkeyrate}
\end{equation}

\subsection{Transfer matrix determination}

To determine the transfer matrix of the proposed QKD system, classical back-projection is implemented (see Results). This means, that Alice uses the VBD as VBE and Bob measures sent states by the VBD or idealized complex conjugated MUB states (see Results). Numerically, we determine probability values in Eq.~(\ref{eq:TranferMatrix}) by calculating the scalar product of sent and measured states (electric fields are represented by matrices) and performing a 2D Fourier transformation (2D FFT). Experimentally, the state sent by Alice would first be collected by a high-numerical-aperture (high-NA) lens or objective and imaged onto the mirror in the NC plane. Next, the light field is imaged onto the decoding plane. For decoding, idealized $\langle \Psi |$ and $\langle \Phi |$ can be implemented by using a combination of $q$- \,\cite{Marrucci2006optical} and wave plates\,\cite{Milione2015, Nape2018}; subsequently, a lens performs the required Fourier transformation, such that, in the lens focus observed on a detector, the on-axis intensity value is proportional to the probability value ($=$ matrix component of $T$). Taking advantage of the 3D polarized MUB states, Bob would use the VBD which is placed in the image plane of Alice's VBE, i.e. the NC plane. The VBD is connected to a single-mode fiber (see Fig.~\ref{fig:Concept}a), guiding its output to the detector (APD).


\begin{thebibliography}{52}%
\makeatletter
\providecommand \@ifxundefined [1]{%
 \@ifx{#1\undefined}
}%
\providecommand \@ifnum [1]{%
 \ifnum #1\expandafter \@firstoftwo
 \else \expandafter \@secondoftwo
 \fi
}%
\providecommand \@ifx [1]{%
 \ifx #1\expandafter \@firstoftwo
 \else \expandafter \@secondoftwo
 \fi
}%
\providecommand \natexlab [1]{#1}%
\providecommand \enquote  [1]{``#1''}%
\providecommand \bibnamefont  [1]{#1}%
\providecommand \bibfnamefont [1]{#1}%
\providecommand \citenamefont [1]{#1}%
\providecommand \href@noop [0]{\@secondoftwo}%
\providecommand \href [0]{\begingroup \@sanitize@url \@href}%
\providecommand \@href[1]{\@@startlink{#1}\@@href}%
\providecommand \@@href[1]{\endgroup#1\@@endlink}%
\providecommand \@sanitize@url [0]{\catcode `\\12\catcode `\$12\catcode
  `\&12\catcode `\#12\catcode `\^12\catcode `\_12\catcode `\%12\relax}%
\providecommand \@@startlink[1]{}%
\providecommand \@@endlink[0]{}%
\providecommand \url  [0]{\begingroup\@sanitize@url \@url }%
\providecommand \@url [1]{\endgroup\@href {#1}{\urlprefix }}%
\providecommand \urlprefix  [0]{URL }%
\providecommand \Eprint [0]{\href }%
\providecommand \doibase [0]{https://doi.org/}%
\providecommand \selectlanguage [0]{\@gobble}%
\providecommand \bibinfo  [0]{\@secondoftwo}%
\providecommand \bibfield  [0]{\@secondoftwo}%
\providecommand \translation [1]{[#1]}%
\providecommand \BibitemOpen [0]{}%
\providecommand \bibitemStop [0]{}%
\providecommand \bibitemNoStop [0]{.\EOS\space}%
\providecommand \EOS [0]{\spacefactor3000\relax}%
\providecommand \BibitemShut  [1]{\csname bibitem#1\endcsname}%
\let\auto@bib@innerbib\@empty
\bibitem [{\citenamefont {O'Brien}\ \emph {et~al.}(2009)\citenamefont
  {O'Brien}, \citenamefont {Furusawa},\ and\ \citenamefont
  {Vu{\v{c}}kovi{\'c}}}]{OBrien2009photonic}%
  \BibitemOpen
  \bibfield  {author} {\bibinfo {author} {\bibfnamefont {J.~L.}\ \bibnamefont
  {O'Brien}}, \bibinfo {author} {\bibfnamefont {A.}~\bibnamefont {Furusawa}},\
  and\ \bibinfo {author} {\bibfnamefont {J.}~\bibnamefont
  {Vu{\v{c}}kovi{\'c}}},\ }\bibfield  {title} {\bibinfo {title} {Photonic
  quantum technologies},\ }\href@noop {} {\bibfield  {journal} {\bibinfo
  {journal} {Nature Photonics}\ }\textbf {\bibinfo {volume} {3}},\ \bibinfo
  {pages} {687} (\bibinfo {year} {2009})}\BibitemShut {NoStop}%
\bibitem [{\citenamefont {Shor}\ and\ \citenamefont
  {Preskill}(2000)}]{shor2000simple}%
  \BibitemOpen
  \bibfield  {author} {\bibinfo {author} {\bibfnamefont {P.~W.}\ \bibnamefont
  {Shor}}\ and\ \bibinfo {author} {\bibfnamefont {J.}~\bibnamefont
  {Preskill}},\ }\bibfield  {title} {\bibinfo {title} {Simple proof of security
  of the bb84 quantum key distribution protocol},\ }\href@noop {} {\bibfield
  {journal} {\bibinfo  {journal} {Phys. Rev. Lett.}\ }\textbf {\bibinfo
  {volume} {85}},\ \bibinfo {pages} {441} (\bibinfo {year} {2000})}\BibitemShut
  {NoStop}%
\bibitem [{\citenamefont {Bennett}\ and\ \citenamefont
  {Brassard}(1984)}]{Bennett1984}%
  \BibitemOpen
  \bibfield  {author} {\bibinfo {author} {\bibfnamefont {C.~H.}\ \bibnamefont
  {Bennett}}\ and\ \bibinfo {author} {\bibfnamefont {G.}~\bibnamefont
  {Brassard}},\ }\bibfield  {title} {\bibinfo {title} {Quantum cryptography:
  Public key distribution and coin tossing},\ }\href@noop {} {\bibfield
  {journal} {\bibinfo  {journal} {Proc. IEEE Int. Conf. on Comput. Syst. Signal
  Process. Bangalore, India}\ ,\ \bibinfo {pages} {175}} (\bibinfo {year}
  {1984})}\BibitemShut {NoStop}%
\bibitem [{\citenamefont {Liao}\ \emph {et~al.}(2017)\citenamefont {Liao},
  \citenamefont {Cai}, \citenamefont {Liu}, \citenamefont {Zhang},
  \citenamefont {Li}, \citenamefont {Ren}, \citenamefont {Yin}, \citenamefont
  {Shen}, \citenamefont {Cao}, \citenamefont {Li}, \citenamefont {Li},
  \citenamefont {Chen}, \citenamefont {Sun}, \citenamefont {Jia}, \citenamefont
  {Wu}, \citenamefont {Jiang}, \citenamefont {Wang}, \citenamefont {Huang},
  \citenamefont {Wang}, \citenamefont {Zhou}, \citenamefont {Deng},
  \citenamefont {Xi}, \citenamefont {Ma}, \citenamefont {Hu}, \citenamefont
  {Zhang}, \citenamefont {Chen}, \citenamefont {Liu}, \citenamefont {Wang},
  \citenamefont {Zhu}, \citenamefont {Lu}, \citenamefont {Shu}, \citenamefont
  {Peng}, \citenamefont {Wang},\ and\ \citenamefont {Pan}}]{liao2017satellite}%
  \BibitemOpen
  \bibfield  {author} {\bibinfo {author} {\bibfnamefont {S.-K.}\ \bibnamefont
  {Liao}}, \bibinfo {author} {\bibfnamefont {W.-Q.}\ \bibnamefont {Cai}},
  \bibinfo {author} {\bibfnamefont {W.-Y.}\ \bibnamefont {Liu}}, \bibinfo
  {author} {\bibfnamefont {L.}~\bibnamefont {Zhang}}, \bibinfo {author}
  {\bibfnamefont {Y.}~\bibnamefont {Li}}, \bibinfo {author} {\bibfnamefont
  {J.-G.}\ \bibnamefont {Ren}}, \bibinfo {author} {\bibfnamefont
  {J.}~\bibnamefont {Yin}}, \bibinfo {author} {\bibfnamefont {Q.}~\bibnamefont
  {Shen}}, \bibinfo {author} {\bibfnamefont {Y.}~\bibnamefont {Cao}}, \bibinfo
  {author} {\bibfnamefont {Z.-P.}\ \bibnamefont {Li}}, \bibinfo {author}
  {\bibfnamefont {F.-Z.}\ \bibnamefont {Li}}, \bibinfo {author} {\bibfnamefont
  {X.-W.}\ \bibnamefont {Chen}}, \bibinfo {author} {\bibfnamefont {L.-H.}\
  \bibnamefont {Sun}}, \bibinfo {author} {\bibfnamefont {J.-J.}\ \bibnamefont
  {Jia}}, \bibinfo {author} {\bibfnamefont {J.-C.}\ \bibnamefont {Wu}},
  \bibinfo {author} {\bibfnamefont {X.-J.}\ \bibnamefont {Jiang}}, \bibinfo
  {author} {\bibfnamefont {J.-F.}\ \bibnamefont {Wang}}, \bibinfo {author}
  {\bibfnamefont {Y.-M.}\ \bibnamefont {Huang}}, \bibinfo {author}
  {\bibfnamefont {Q.}~\bibnamefont {Wang}}, \bibinfo {author} {\bibfnamefont
  {Y.-L.}\ \bibnamefont {Zhou}}, \bibinfo {author} {\bibfnamefont
  {L.}~\bibnamefont {Deng}}, \bibinfo {author} {\bibfnamefont {T.}~\bibnamefont
  {Xi}}, \bibinfo {author} {\bibfnamefont {L.}~\bibnamefont {Ma}}, \bibinfo
  {author} {\bibfnamefont {T.}~\bibnamefont {Hu}}, \bibinfo {author}
  {\bibfnamefont {Q.}~\bibnamefont {Zhang}}, \bibinfo {author} {\bibfnamefont
  {Y.-A.}\ \bibnamefont {Chen}}, \bibinfo {author} {\bibfnamefont {N.-L.}\
  \bibnamefont {Liu}}, \bibinfo {author} {\bibfnamefont {X.-B.}\ \bibnamefont
  {Wang}}, \bibinfo {author} {\bibfnamefont {Z.-C.}\ \bibnamefont {Zhu}},
  \bibinfo {author} {\bibfnamefont {C.-Y.}\ \bibnamefont {Lu}}, \bibinfo
  {author} {\bibfnamefont {R.}~\bibnamefont {Shu}}, \bibinfo {author}
  {\bibfnamefont {C.-Z.}\ \bibnamefont {Peng}}, \bibinfo {author}
  {\bibfnamefont {J.-Y.}\ \bibnamefont {Wang}},\ and\ \bibinfo {author}
  {\bibfnamefont {J.-W.}\ \bibnamefont {Pan}},\ }\bibfield  {title} {\bibinfo
  {title} {Satellite-to-ground quantum key distribution},\ }\href@noop {}
  {\bibfield  {journal} {\bibinfo  {journal} {Nature}\ }\textbf {\bibinfo
  {volume} {549}},\ \bibinfo {pages} {43} (\bibinfo {year} {2017})}\BibitemShut
  {NoStop}%
\bibitem [{\citenamefont {Boaron}\ \emph {et~al.}(2018)\citenamefont {Boaron},
  \citenamefont {Boso}, \citenamefont {Rusca}, \citenamefont {Vulliez},
  \citenamefont {Autebert}, \citenamefont {Caloz}, \citenamefont {Perrenoud},
  \citenamefont {Gras}, \citenamefont {Bussi\`eres}, \citenamefont {Li},
  \citenamefont {Nolan}, \citenamefont {Martin},\ and\ \citenamefont
  {Zbinden}}]{Boaron2018}%
  \BibitemOpen
  \bibfield  {author} {\bibinfo {author} {\bibfnamefont {A.}~\bibnamefont
  {Boaron}}, \bibinfo {author} {\bibfnamefont {G.}~\bibnamefont {Boso}},
  \bibinfo {author} {\bibfnamefont {D.}~\bibnamefont {Rusca}}, \bibinfo
  {author} {\bibfnamefont {C.}~\bibnamefont {Vulliez}}, \bibinfo {author}
  {\bibfnamefont {C.}~\bibnamefont {Autebert}}, \bibinfo {author}
  {\bibfnamefont {M.}~\bibnamefont {Caloz}}, \bibinfo {author} {\bibfnamefont
  {M.}~\bibnamefont {Perrenoud}}, \bibinfo {author} {\bibfnamefont
  {G.}~\bibnamefont {Gras}}, \bibinfo {author} {\bibfnamefont {F.}~\bibnamefont
  {Bussi\`eres}}, \bibinfo {author} {\bibfnamefont {M.-J.}\ \bibnamefont {Li}},
  \bibinfo {author} {\bibfnamefont {D.}~\bibnamefont {Nolan}}, \bibinfo
  {author} {\bibfnamefont {A.}~\bibnamefont {Martin}},\ and\ \bibinfo {author}
  {\bibfnamefont {H.}~\bibnamefont {Zbinden}},\ }\bibfield  {title} {\bibinfo
  {title} {Secure quantum key distribution over 421 km of optical fiber},\
  }\href {https://doi.org/10.1103/PhysRevLett.121.190502} {\bibfield  {journal}
  {\bibinfo  {journal} {Phys. Rev. Lett.}\ }\textbf {\bibinfo {volume} {121}},\
  \bibinfo {pages} {190502} (\bibinfo {year} {2018})}\BibitemShut {NoStop}%
\bibitem [{\citenamefont {Ding}\ \emph {et~al.}(2017)\citenamefont {Ding},
  \citenamefont {Bacco}, \citenamefont {Dalgaard}, \citenamefont {Cai},
  \citenamefont {Zhou}, \citenamefont {Rottwitt},\ and\ \citenamefont
  {Oxenl{\o}we}}]{Ding2017}%
  \BibitemOpen
  \bibfield  {author} {\bibinfo {author} {\bibfnamefont {Y.}~\bibnamefont
  {Ding}}, \bibinfo {author} {\bibfnamefont {D.}~\bibnamefont {Bacco}},
  \bibinfo {author} {\bibfnamefont {K.}~\bibnamefont {Dalgaard}}, \bibinfo
  {author} {\bibfnamefont {X.}~\bibnamefont {Cai}}, \bibinfo {author}
  {\bibfnamefont {X.}~\bibnamefont {Zhou}}, \bibinfo {author} {\bibfnamefont
  {K.}~\bibnamefont {Rottwitt}},\ and\ \bibinfo {author} {\bibfnamefont
  {L.~K.}\ \bibnamefont {Oxenl{\o}we}},\ }\bibfield  {title} {\bibinfo {title}
  {High-dimensional quantum key distribution based on multicore fiber using
  silicon photonic integrated circuits},\ }\href@noop {} {\bibfield  {journal}
  {\bibinfo  {journal} {npj Quantum Information}\ }\textbf {\bibinfo {volume}
  {3}},\ \bibinfo {pages} {25} (\bibinfo {year} {2017})}\BibitemShut {NoStop}%
\bibitem [{\citenamefont {Chen}\ \emph {et~al.}(2022)\citenamefont {Chen},
  \citenamefont {Zhang}, \citenamefont {Liu}, \citenamefont {Jiang},
  \citenamefont {Zhao}, \citenamefont {Zhang}, \citenamefont {Chen},
  \citenamefont {Li}, \citenamefont {You}, \citenamefont {Wang} \emph
  {et~al.}}]{Chen2022quantum}%
  \BibitemOpen
  \bibfield  {author} {\bibinfo {author} {\bibfnamefont {J.-P.}\ \bibnamefont
  {Chen}}, \bibinfo {author} {\bibfnamefont {C.}~\bibnamefont {Zhang}},
  \bibinfo {author} {\bibfnamefont {Y.}~\bibnamefont {Liu}}, \bibinfo {author}
  {\bibfnamefont {C.}~\bibnamefont {Jiang}}, \bibinfo {author} {\bibfnamefont
  {D.-F.}\ \bibnamefont {Zhao}}, \bibinfo {author} {\bibfnamefont {W.-J.}\
  \bibnamefont {Zhang}}, \bibinfo {author} {\bibfnamefont {F.-X.}\ \bibnamefont
  {Chen}}, \bibinfo {author} {\bibfnamefont {H.}~\bibnamefont {Li}}, \bibinfo
  {author} {\bibfnamefont {L.-X.}\ \bibnamefont {You}}, \bibinfo {author}
  {\bibfnamefont {Z.}~\bibnamefont {Wang}}, \emph {et~al.},\ }\bibfield
  {title} {\bibinfo {title} {Quantum key distribution over 658 km fiber with
  distributed vibration sensing},\ }\href@noop {} {\bibfield  {journal}
  {\bibinfo  {journal} {Phys. Rev. Lett.}\ }\textbf {\bibinfo {volume} {128}},\
  \bibinfo {pages} {180502} (\bibinfo {year} {2022})}\BibitemShut {NoStop}%
\bibitem [{\citenamefont {Bechmann-Pasquinucci}\ and\ \citenamefont
  {Tittel}(2000)}]{bechmann2000quantum}%
  \BibitemOpen
  \bibfield  {author} {\bibinfo {author} {\bibfnamefont {H.}~\bibnamefont
  {Bechmann-Pasquinucci}}\ and\ \bibinfo {author} {\bibfnamefont
  {W.}~\bibnamefont {Tittel}},\ }\bibfield  {title} {\bibinfo {title} {Quantum
  cryptography using larger alphabets},\ }\href@noop {} {\bibfield  {journal}
  {\bibinfo  {journal} {Phys Rev A}\ }\textbf {\bibinfo {volume} {61}},\
  \bibinfo {pages} {062308} (\bibinfo {year} {2000})}\BibitemShut {NoStop}%
\bibitem [{\citenamefont {Ali-Khan}\ \emph {et~al.}(2007)\citenamefont
  {Ali-Khan}, \citenamefont {Broadbent},\ and\ \citenamefont
  {Howell}}]{ali2007large}%
  \BibitemOpen
  \bibfield  {author} {\bibinfo {author} {\bibfnamefont {I.}~\bibnamefont
  {Ali-Khan}}, \bibinfo {author} {\bibfnamefont {C.~J.}\ \bibnamefont
  {Broadbent}},\ and\ \bibinfo {author} {\bibfnamefont {J.~C.}\ \bibnamefont
  {Howell}},\ }\bibfield  {title} {\bibinfo {title} {Large-alphabet quantum key
  distribution using energy-time entangled bipartite states},\ }\href@noop {}
  {\bibfield  {journal} {\bibinfo  {journal} {Phys Rev Lett}\ }\textbf
  {\bibinfo {volume} {98}},\ \bibinfo {pages} {060503} (\bibinfo {year}
  {2007})}\BibitemShut {NoStop}%
\bibitem [{\citenamefont {Cerf}\ \emph {et~al.}(2002)\citenamefont {Cerf},
  \citenamefont {Bourennane}, \citenamefont {Karlsson},\ and\ \citenamefont
  {Gisin}}]{cerf2002security}%
  \BibitemOpen
  \bibfield  {author} {\bibinfo {author} {\bibfnamefont {N.~J.}\ \bibnamefont
  {Cerf}}, \bibinfo {author} {\bibfnamefont {M.}~\bibnamefont {Bourennane}},
  \bibinfo {author} {\bibfnamefont {A.}~\bibnamefont {Karlsson}},\ and\
  \bibinfo {author} {\bibfnamefont {N.}~\bibnamefont {Gisin}},\ }\bibfield
  {title} {\bibinfo {title} {Security of quantum key distribution using d-level
  systems},\ }\href@noop {} {\bibfield  {journal} {\bibinfo  {journal} {Phys
  Rev Lett}\ }\textbf {\bibinfo {volume} {88}},\ \bibinfo {pages} {127902}
  (\bibinfo {year} {2002})}\BibitemShut {NoStop}%
\bibitem [{\citenamefont {Erhard}\ \emph {et~al.}(2020)\citenamefont {Erhard},
  \citenamefont {Krenn},\ and\ \citenamefont
  {Zeilinger}}]{Erhard2020advancesHD}%
  \BibitemOpen
  \bibfield  {author} {\bibinfo {author} {\bibfnamefont {M.}~\bibnamefont
  {Erhard}}, \bibinfo {author} {\bibfnamefont {M.}~\bibnamefont {Krenn}},\ and\
  \bibinfo {author} {\bibfnamefont {A.}~\bibnamefont {Zeilinger}},\ }\bibfield
  {title} {\bibinfo {title} {Advances in high-dimensional quantum
  entanglement},\ }\href@noop {} {\bibfield  {journal} {\bibinfo  {journal}
  {Nature Reviews Physics}\ }\textbf {\bibinfo {volume} {2}},\ \bibinfo {pages}
  {365} (\bibinfo {year} {2020})}\BibitemShut {NoStop}%
\bibitem [{\citenamefont {Otte}\ \emph {et~al.}(2020)\citenamefont {Otte},
  \citenamefont {Nape}, \citenamefont {Rosales-Guzm\'{a}n}, \citenamefont
  {Denz}, \citenamefont {Forbes},\ and\ \citenamefont
  {Ndagano}}]{Otte2020Tutorial}%
  \BibitemOpen
  \bibfield  {author} {\bibinfo {author} {\bibfnamefont {E.}~\bibnamefont
  {Otte}}, \bibinfo {author} {\bibfnamefont {I.}~\bibnamefont {Nape}}, \bibinfo
  {author} {\bibfnamefont {C.}~\bibnamefont {Rosales-Guzm\'{a}n}}, \bibinfo
  {author} {\bibfnamefont {C.}~\bibnamefont {Denz}}, \bibinfo {author}
  {\bibfnamefont {A.}~\bibnamefont {Forbes}},\ and\ \bibinfo {author}
  {\bibfnamefont {B.}~\bibnamefont {Ndagano}},\ }\bibfield  {title} {\bibinfo
  {title} {High-dimensional cryptography with spatial modes of light:
  tutorial},\ }\href {https://doi.org/10.1364/JOSAB.399290} {\bibfield
  {journal} {\bibinfo  {journal} {J. Opt. Soc. Am. B}\ }\textbf {\bibinfo
  {volume} {37}},\ \bibinfo {pages} {A309} (\bibinfo {year}
  {2020})}\BibitemShut {NoStop}%
\bibitem [{\citenamefont {Erhard}\ \emph {et~al.}(2018)\citenamefont {Erhard},
  \citenamefont {Fickler}, \citenamefont {Krenn},\ and\ \citenamefont
  {Zeilinger}}]{LSA7}%
  \BibitemOpen
  \bibfield  {author} {\bibinfo {author} {\bibfnamefont {M.}~\bibnamefont
  {Erhard}}, \bibinfo {author} {\bibfnamefont {R.}~\bibnamefont {Fickler}},
  \bibinfo {author} {\bibfnamefont {M.}~\bibnamefont {Krenn}},\ and\ \bibinfo
  {author} {\bibfnamefont {A.}~\bibnamefont {Zeilinger}},\ }\bibfield  {title}
  {\bibinfo {title} {Twisted photons: new quantum perspectives in high
  dimensions},\ }\href {https://doi.org/10.1038/lsa.2017.146} {\bibfield
  {journal} {\bibinfo  {journal} {Light: Science and Applications}\ }\textbf
  {\bibinfo {volume} {7}},\ \bibinfo {pages} {17146} (\bibinfo {year}
  {2018})}\BibitemShut {NoStop}%
\bibitem [{\citenamefont {Ndagano}\ \emph {et~al.}(2018)\citenamefont
  {Ndagano}, \citenamefont {Nape}, \citenamefont {Cox}, \citenamefont
  {Rosales-Guzman},\ and\ \citenamefont {Forbes}}]{Ndagano2018}%
  \BibitemOpen
  \bibfield  {author} {\bibinfo {author} {\bibfnamefont {B.}~\bibnamefont
  {Ndagano}}, \bibinfo {author} {\bibfnamefont {I.}~\bibnamefont {Nape}},
  \bibinfo {author} {\bibfnamefont {M.~A.}\ \bibnamefont {Cox}}, \bibinfo
  {author} {\bibfnamefont {C.}~\bibnamefont {Rosales-Guzman}},\ and\ \bibinfo
  {author} {\bibfnamefont {A.}~\bibnamefont {Forbes}},\ }\bibfield  {title}
  {\bibinfo {title} {{Creation and detection of vector vortex modes for
  classical and quantum communication}},\ }\href
  {https://doi.org/10.1109/JLT.2017.2766760} {\bibfield  {journal} {\bibinfo
  {journal} {Journal of Lightwave Technology}\ }\textbf {\bibinfo {volume}
  {36}},\ \bibinfo {pages} {292} (\bibinfo {year} {2018})}\BibitemShut
  {NoStop}%
\bibitem [{\citenamefont {Mirhosseini}\ \emph {et~al.}(2015)\citenamefont
  {Mirhosseini}, \citenamefont {Maga{\~n}a-Loaiza}, \citenamefont {O'Sullivan},
  \citenamefont {Rodenburg}, \citenamefont {Malik}, \citenamefont {Lavery},
  \citenamefont {Padgett}, \citenamefont {Gauthier},\ and\ \citenamefont
  {Boyd}}]{mirhosseini2015high}%
  \BibitemOpen
  \bibfield  {author} {\bibinfo {author} {\bibfnamefont {M.}~\bibnamefont
  {Mirhosseini}}, \bibinfo {author} {\bibfnamefont {O.~S.}\ \bibnamefont
  {Maga{\~n}a-Loaiza}}, \bibinfo {author} {\bibfnamefont {M.~N.}\ \bibnamefont
  {O'Sullivan}}, \bibinfo {author} {\bibfnamefont {B.}~\bibnamefont
  {Rodenburg}}, \bibinfo {author} {\bibfnamefont {M.}~\bibnamefont {Malik}},
  \bibinfo {author} {\bibfnamefont {M.~P.}\ \bibnamefont {Lavery}}, \bibinfo
  {author} {\bibfnamefont {M.~J.}\ \bibnamefont {Padgett}}, \bibinfo {author}
  {\bibfnamefont {D.~J.}\ \bibnamefont {Gauthier}},\ and\ \bibinfo {author}
  {\bibfnamefont {R.~W.}\ \bibnamefont {Boyd}},\ }\bibfield  {title} {\bibinfo
  {title} {High-dimensional quantum cryptography with twisted light},\
  }\href@noop {} {\bibfield  {journal} {\bibinfo  {journal} {New J Phys}\
  }\textbf {\bibinfo {volume} {17}},\ \bibinfo {pages} {033033} (\bibinfo
  {year} {2015})}\BibitemShut {NoStop}%
\bibitem [{\citenamefont {Hufnagel}\ \emph {et~al.}(2020)\citenamefont
  {Hufnagel}, \citenamefont {Sit}, \citenamefont {Bouchard}, \citenamefont
  {Zhang}, \citenamefont {England}, \citenamefont {Heshami}, \citenamefont
  {Sussman},\ and\ \citenamefont {Karimi}}]{Hufnagel2020}%
  \BibitemOpen
  \bibfield  {author} {\bibinfo {author} {\bibfnamefont {F.}~\bibnamefont
  {Hufnagel}}, \bibinfo {author} {\bibfnamefont {A.}~\bibnamefont {Sit}},
  \bibinfo {author} {\bibfnamefont {F.}~\bibnamefont {Bouchard}}, \bibinfo
  {author} {\bibfnamefont {Y.}~\bibnamefont {Zhang}}, \bibinfo {author}
  {\bibfnamefont {D.}~\bibnamefont {England}}, \bibinfo {author} {\bibfnamefont
  {K.}~\bibnamefont {Heshami}}, \bibinfo {author} {\bibfnamefont {B.~J.}\
  \bibnamefont {Sussman}},\ and\ \bibinfo {author} {\bibfnamefont
  {E.}~\bibnamefont {Karimi}},\ }\bibfield  {title} {\bibinfo {title}
  {Underwater quantum communication over a 30-meter flume tank},\ }\href@noop
  {} {\bibfield  {journal} {\bibinfo  {journal} {arXiv 2004.04821}\ } (\bibinfo
  {year} {2020})}\BibitemShut {NoStop}%
\bibitem [{\citenamefont {Nape}\ \emph {et~al.}(2018)\citenamefont {Nape},
  \citenamefont {Otte}, \citenamefont {Vall\'{e}s}, \citenamefont
  {Rosales-Guzm\'{a}n}, \citenamefont {Cardano}, \citenamefont {Denz},\ and\
  \citenamefont {Forbes}}]{Nape2018}%
  \BibitemOpen
  \bibfield  {author} {\bibinfo {author} {\bibfnamefont {I.}~\bibnamefont
  {Nape}}, \bibinfo {author} {\bibfnamefont {E.}~\bibnamefont {Otte}}, \bibinfo
  {author} {\bibfnamefont {A.}~\bibnamefont {Vall\'{e}s}}, \bibinfo {author}
  {\bibfnamefont {C.}~\bibnamefont {Rosales-Guzm\'{a}n}}, \bibinfo {author}
  {\bibfnamefont {F.}~\bibnamefont {Cardano}}, \bibinfo {author} {\bibfnamefont
  {C.}~\bibnamefont {Denz}},\ and\ \bibinfo {author} {\bibfnamefont
  {A.}~\bibnamefont {Forbes}},\ }\bibfield  {title} {\bibinfo {title}
  {Self-healing high-dimensional quantum key distribution using hybrid
  spin-orbit bessel states},\ }\href {https://doi.org/10.1364/OE.26.026946}
  {\bibfield  {journal} {\bibinfo  {journal} {Opt Express}\ }\textbf {\bibinfo
  {volume} {26}},\ \bibinfo {pages} {26946} (\bibinfo {year}
  {2018})}\BibitemShut {NoStop}%
\bibitem [{\citenamefont {Sit}\ \emph {et~al.}(2017)\citenamefont {Sit},
  \citenamefont {Bouchard}, \citenamefont {Fickler}, \citenamefont
  {Gagnon-Bischoff}, \citenamefont {Larocque}, \citenamefont {Heshami},
  \citenamefont {Elser}, \citenamefont {Peuntinger}, \citenamefont
  {G{\"{u}}nthner}, \citenamefont {Heim}, \citenamefont {Marquardt},
  \citenamefont {Leuchs}, \citenamefont {Boyd},\ and\ \citenamefont
  {Karimi}}]{Sit2017}%
  \BibitemOpen
  \bibfield  {author} {\bibinfo {author} {\bibfnamefont {A.}~\bibnamefont
  {Sit}}, \bibinfo {author} {\bibfnamefont {F.}~\bibnamefont {Bouchard}},
  \bibinfo {author} {\bibfnamefont {R.}~\bibnamefont {Fickler}}, \bibinfo
  {author} {\bibfnamefont {J.}~\bibnamefont {Gagnon-Bischoff}}, \bibinfo
  {author} {\bibfnamefont {H.}~\bibnamefont {Larocque}}, \bibinfo {author}
  {\bibfnamefont {K.}~\bibnamefont {Heshami}}, \bibinfo {author} {\bibfnamefont
  {D.}~\bibnamefont {Elser}}, \bibinfo {author} {\bibfnamefont
  {C.}~\bibnamefont {Peuntinger}}, \bibinfo {author} {\bibfnamefont
  {K.}~\bibnamefont {G{\"{u}}nthner}}, \bibinfo {author} {\bibfnamefont
  {B.}~\bibnamefont {Heim}}, \bibinfo {author} {\bibfnamefont {C.}~\bibnamefont
  {Marquardt}}, \bibinfo {author} {\bibfnamefont {G.}~\bibnamefont {Leuchs}},
  \bibinfo {author} {\bibfnamefont {R.~W.}\ \bibnamefont {Boyd}},\ and\
  \bibinfo {author} {\bibfnamefont {E.}~\bibnamefont {Karimi}},\ }\bibfield
  {title} {\bibinfo {title} {{High-dimensional intracity quantum cryptography
  with structured photons}},\ }\href {https://doi.org/10.1364/OPTICA.4.001006}
  {\bibfield  {journal} {\bibinfo  {journal} {Optica}\ }\textbf {\bibinfo
  {volume} {4}},\ \bibinfo {pages} {1006} (\bibinfo {year} {2017})},\ \Eprint
  {https://arxiv.org/abs/1612.05195} {arXiv:1612.05195} \BibitemShut {NoStop}%
\bibitem [{\citenamefont {Cozzolino}\ \emph {et~al.}(2019)\citenamefont
  {Cozzolino}, \citenamefont {Bacco}, \citenamefont {Da~Lio}, \citenamefont
  {Ingerslev}, \citenamefont {Ding}, \citenamefont {Dalgaard}, \citenamefont
  {Kristensen}, \citenamefont {Galili}, \citenamefont {Rottwitt}, \citenamefont
  {Ramachandran},\ and\ \citenamefont {Oxenl\o{}we}}]{cozzolino2019orbital}%
  \BibitemOpen
  \bibfield  {author} {\bibinfo {author} {\bibfnamefont {D.}~\bibnamefont
  {Cozzolino}}, \bibinfo {author} {\bibfnamefont {D.}~\bibnamefont {Bacco}},
  \bibinfo {author} {\bibfnamefont {B.}~\bibnamefont {Da~Lio}}, \bibinfo
  {author} {\bibfnamefont {K.}~\bibnamefont {Ingerslev}}, \bibinfo {author}
  {\bibfnamefont {Y.}~\bibnamefont {Ding}}, \bibinfo {author} {\bibfnamefont
  {K.}~\bibnamefont {Dalgaard}}, \bibinfo {author} {\bibfnamefont
  {P.}~\bibnamefont {Kristensen}}, \bibinfo {author} {\bibfnamefont
  {M.}~\bibnamefont {Galili}}, \bibinfo {author} {\bibfnamefont
  {K.}~\bibnamefont {Rottwitt}}, \bibinfo {author} {\bibfnamefont
  {S.}~\bibnamefont {Ramachandran}},\ and\ \bibinfo {author} {\bibfnamefont
  {L.~K.}\ \bibnamefont {Oxenl\o{}we}},\ }\bibfield  {title} {\bibinfo {title}
  {Orbital angular momentum states enabling fiber-based high-dimensional
  quantum communication},\ }\href
  {https://doi.org/10.1103/PhysRevApplied.11.064058} {\bibfield  {journal}
  {\bibinfo  {journal} {Phys. Rev. Applied}\ }\textbf {\bibinfo {volume}
  {11}},\ \bibinfo {pages} {064058} (\bibinfo {year} {2019})}\BibitemShut
  {NoStop}%
\bibitem [{\citenamefont {Bouchard}\ \emph {et~al.}(2018)\citenamefont
  {Bouchard}, \citenamefont {Sit}, \citenamefont {Hufnagel}, \citenamefont
  {Abbas}, \citenamefont {Zhang}, \citenamefont {Heshami}, \citenamefont
  {Fickler}, \citenamefont {Marquardt}, \citenamefont {Leuchs}, \citenamefont
  {Boyd},\ and\ \citenamefont {Karimi}}]{bouchard2018underwater}%
  \BibitemOpen
  \bibfield  {author} {\bibinfo {author} {\bibfnamefont {F.}~\bibnamefont
  {Bouchard}}, \bibinfo {author} {\bibfnamefont {A.}~\bibnamefont {Sit}},
  \bibinfo {author} {\bibfnamefont {F.}~\bibnamefont {Hufnagel}}, \bibinfo
  {author} {\bibfnamefont {A.}~\bibnamefont {Abbas}}, \bibinfo {author}
  {\bibfnamefont {Y.}~\bibnamefont {Zhang}}, \bibinfo {author} {\bibfnamefont
  {K.}~\bibnamefont {Heshami}}, \bibinfo {author} {\bibfnamefont
  {R.}~\bibnamefont {Fickler}}, \bibinfo {author} {\bibfnamefont
  {C.}~\bibnamefont {Marquardt}}, \bibinfo {author} {\bibfnamefont
  {G.}~\bibnamefont {Leuchs}}, \bibinfo {author} {\bibfnamefont {R.~w.}\
  \bibnamefont {Boyd}},\ and\ \bibinfo {author} {\bibfnamefont
  {E.}~\bibnamefont {Karimi}},\ }\bibfield  {title} {\bibinfo {title} {{Quantum
  cryptography with twisted photons through an outdoor underwater channel}},\
  }\href {https://doi.org/10.1364/oe.26.022563} {\bibfield  {journal} {\bibinfo
   {journal} {Optics Express}\ }\textbf {\bibinfo {volume} {26}},\ \bibinfo
  {pages} {22563} (\bibinfo {year} {2018})}\BibitemShut {NoStop}%
\bibitem [{\citenamefont {Berkhout}\ \emph {et~al.}(2010)\citenamefont
  {Berkhout}, \citenamefont {Lavery}, \citenamefont {Courtial}, \citenamefont
  {Beijersbergen},\ and\ \citenamefont {Padgett}}]{berkhout2010efficient}%
  \BibitemOpen
  \bibfield  {author} {\bibinfo {author} {\bibfnamefont {G.~C.}\ \bibnamefont
  {Berkhout}}, \bibinfo {author} {\bibfnamefont {M.~P.}\ \bibnamefont
  {Lavery}}, \bibinfo {author} {\bibfnamefont {J.}~\bibnamefont {Courtial}},
  \bibinfo {author} {\bibfnamefont {M.~W.}\ \bibnamefont {Beijersbergen}},\
  and\ \bibinfo {author} {\bibfnamefont {M.~J.}\ \bibnamefont {Padgett}},\
  }\bibfield  {title} {\bibinfo {title} {Efficient sorting of orbital angular
  momentum states of light},\ }\href@noop {} {\bibfield  {journal} {\bibinfo
  {journal} {Phys Rev Lett}\ }\textbf {\bibinfo {volume} {105}},\ \bibinfo
  {pages} {153601} (\bibinfo {year} {2010})}\BibitemShut {NoStop}%
\bibitem [{\citenamefont {Fickler}\ \emph {et~al.}(2014)\citenamefont
  {Fickler}, \citenamefont {Lapkiewicz}, \citenamefont {Ramelow},\ and\
  \citenamefont {Zeilinger}}]{Fickler2014quantum}%
  \BibitemOpen
  \bibfield  {author} {\bibinfo {author} {\bibfnamefont {R.}~\bibnamefont
  {Fickler}}, \bibinfo {author} {\bibfnamefont {R.}~\bibnamefont {Lapkiewicz}},
  \bibinfo {author} {\bibfnamefont {S.}~\bibnamefont {Ramelow}},\ and\ \bibinfo
  {author} {\bibfnamefont {A.}~\bibnamefont {Zeilinger}},\ }\bibfield  {title}
  {\bibinfo {title} {Quantum entanglement of complex photon polarization
  patterns in vector beams},\ }\href@noop {} {\bibfield  {journal} {\bibinfo
  {journal} {Physical Review A}\ }\textbf {\bibinfo {volume} {89}},\ \bibinfo
  {pages} {060301} (\bibinfo {year} {2014})}\BibitemShut {NoStop}%
\bibitem [{\citenamefont {Milione}\ \emph {et~al.}(2015)\citenamefont
  {Milione}, \citenamefont {Lavery}, \citenamefont {Huang}, \citenamefont
  {Ren}, \citenamefont {Xie}, \citenamefont {Nguyen}, \citenamefont {Karimi},
  \citenamefont {Marrucci}, \citenamefont {Nolan}, \citenamefont {Alfano} \emph
  {et~al.}}]{Milione2015}%
  \BibitemOpen
  \bibfield  {author} {\bibinfo {author} {\bibfnamefont {G.}~\bibnamefont
  {Milione}}, \bibinfo {author} {\bibfnamefont {M.~P.}\ \bibnamefont {Lavery}},
  \bibinfo {author} {\bibfnamefont {H.}~\bibnamefont {Huang}}, \bibinfo
  {author} {\bibfnamefont {Y.}~\bibnamefont {Ren}}, \bibinfo {author}
  {\bibfnamefont {G.}~\bibnamefont {Xie}}, \bibinfo {author} {\bibfnamefont
  {T.~A.}\ \bibnamefont {Nguyen}}, \bibinfo {author} {\bibfnamefont
  {E.}~\bibnamefont {Karimi}}, \bibinfo {author} {\bibfnamefont
  {L.}~\bibnamefont {Marrucci}}, \bibinfo {author} {\bibfnamefont {D.~A.}\
  \bibnamefont {Nolan}}, \bibinfo {author} {\bibfnamefont {R.~R.}\ \bibnamefont
  {Alfano}}, \emph {et~al.},\ }\bibfield  {title} {\bibinfo {title} {4$\times$
  20 gbit/s mode division multiplexing over free space using vector modes and a
  q-plate mode (de) multiplexer},\ }\href@noop {} {\bibfield  {journal}
  {\bibinfo  {journal} {Optics letters}\ }\textbf {\bibinfo {volume} {40}},\
  \bibinfo {pages} {1980} (\bibinfo {year} {2015})}\BibitemShut {NoStop}%
\bibitem [{\citenamefont {Forbes}\ \emph {et~al.}(2016)\citenamefont {Forbes},
  \citenamefont {Dudley},\ and\ \citenamefont {McLaren}}]{Forbes2016}%
  \BibitemOpen
  \bibfield  {author} {\bibinfo {author} {\bibfnamefont {A.}~\bibnamefont
  {Forbes}}, \bibinfo {author} {\bibfnamefont {A.}~\bibnamefont {Dudley}},\
  and\ \bibinfo {author} {\bibfnamefont {M.}~\bibnamefont {McLaren}},\
  }\bibfield  {title} {\bibinfo {title} {{Creation and detection of optical
  modes with spatial light modulators}},\ }\href
  {https://doi.org/10.1364/AOP.8.000200} {\bibfield  {journal} {\bibinfo
  {journal} {Advances in Optics and Photonics}\ }\textbf {\bibinfo {volume}
  {8}},\ \bibinfo {pages} {200} (\bibinfo {year} {2016})}\BibitemShut {NoStop}%
\bibitem [{\citenamefont {Ndagano}\ \emph {et~al.}(2017)\citenamefont
  {Ndagano}, \citenamefont {Nape}, \citenamefont {Perez-Garcia}, \citenamefont
  {Scholes}, \citenamefont {Hernandez-Aranda}, \citenamefont {Konrad},
  \citenamefont {Lavery},\ and\ \citenamefont {Forbes}}]{ndagano2017det}%
  \BibitemOpen
  \bibfield  {author} {\bibinfo {author} {\bibfnamefont {B.}~\bibnamefont
  {Ndagano}}, \bibinfo {author} {\bibfnamefont {I.}~\bibnamefont {Nape}},
  \bibinfo {author} {\bibfnamefont {B.}~\bibnamefont {Perez-Garcia}}, \bibinfo
  {author} {\bibfnamefont {S.}~\bibnamefont {Scholes}}, \bibinfo {author}
  {\bibfnamefont {R.~I.}\ \bibnamefont {Hernandez-Aranda}}, \bibinfo {author}
  {\bibfnamefont {T.}~\bibnamefont {Konrad}}, \bibinfo {author} {\bibfnamefont
  {M.~P.}\ \bibnamefont {Lavery}},\ and\ \bibinfo {author} {\bibfnamefont
  {A.}~\bibnamefont {Forbes}},\ }\bibfield  {title} {\bibinfo {title} {A
  deterministic detector for vector vortex states},\ }\href@noop {} {\bibfield
  {journal} {\bibinfo  {journal} {Sci Rep}\ }\textbf {\bibinfo {volume} {7}},\
  \bibinfo {pages} {13882} (\bibinfo {year} {2017})}\BibitemShut {NoStop}%
\bibitem [{\citenamefont {Fontaine}\ \emph {et~al.}(2019)\citenamefont
  {Fontaine}, \citenamefont {Ryf}, \citenamefont {Chen}, \citenamefont
  {Neilson}, \citenamefont {Kim},\ and\ \citenamefont
  {Carpenter}}]{Fontaine2019laguerre}%
  \BibitemOpen
  \bibfield  {author} {\bibinfo {author} {\bibfnamefont {N.~K.}\ \bibnamefont
  {Fontaine}}, \bibinfo {author} {\bibfnamefont {R.}~\bibnamefont {Ryf}},
  \bibinfo {author} {\bibfnamefont {H.}~\bibnamefont {Chen}}, \bibinfo {author}
  {\bibfnamefont {D.~T.}\ \bibnamefont {Neilson}}, \bibinfo {author}
  {\bibfnamefont {K.}~\bibnamefont {Kim}},\ and\ \bibinfo {author}
  {\bibfnamefont {J.}~\bibnamefont {Carpenter}},\ }\bibfield  {title} {\bibinfo
  {title} {Laguerre-gaussian mode sorter},\ }\href@noop {} {\bibfield
  {journal} {\bibinfo  {journal} {Nature communications}\ }\textbf {\bibinfo
  {volume} {10}},\ \bibinfo {pages} {1} (\bibinfo {year} {2019})}\BibitemShut
  {NoStop}%
\bibitem [{\citenamefont {Fickler}\ \emph {et~al.}(2020)\citenamefont
  {Fickler}, \citenamefont {Bouchard}, \citenamefont {Giese}, \citenamefont
  {Grillo}, \citenamefont {Leuchs},\ and\ \citenamefont
  {Karimi}}]{fickler2020full}%
  \BibitemOpen
  \bibfield  {author} {\bibinfo {author} {\bibfnamefont {R.}~\bibnamefont
  {Fickler}}, \bibinfo {author} {\bibfnamefont {F.}~\bibnamefont {Bouchard}},
  \bibinfo {author} {\bibfnamefont {E.}~\bibnamefont {Giese}}, \bibinfo
  {author} {\bibfnamefont {V.}~\bibnamefont {Grillo}}, \bibinfo {author}
  {\bibfnamefont {G.}~\bibnamefont {Leuchs}},\ and\ \bibinfo {author}
  {\bibfnamefont {E.}~\bibnamefont {Karimi}},\ }\bibfield  {title} {\bibinfo
  {title} {Full-field mode sorter using two optimized phase transformations for
  high-dimensional quantum cryptography},\ }\href@noop {} {\bibfield  {journal}
  {\bibinfo  {journal} {Journal of Optics}\ }\textbf {\bibinfo {volume} {22}},\
  \bibinfo {pages} {024001} (\bibinfo {year} {2020})}\BibitemShut {NoStop}%
\bibitem [{\citenamefont {Wang}\ \emph {et~al.}(2020)\citenamefont {Wang},
  \citenamefont {Sciarrino}, \citenamefont {Laing},\ and\ \citenamefont
  {Thompson}}]{Wang2020integrated}%
  \BibitemOpen
  \bibfield  {author} {\bibinfo {author} {\bibfnamefont {J.}~\bibnamefont
  {Wang}}, \bibinfo {author} {\bibfnamefont {F.}~\bibnamefont {Sciarrino}},
  \bibinfo {author} {\bibfnamefont {A.}~\bibnamefont {Laing}},\ and\ \bibinfo
  {author} {\bibfnamefont {M.~G.}\ \bibnamefont {Thompson}},\ }\bibfield
  {title} {\bibinfo {title} {Integrated photonic quantum technologies},\
  }\href@noop {} {\bibfield  {journal} {\bibinfo  {journal} {Nature Photonics}\
  }\textbf {\bibinfo {volume} {14}},\ \bibinfo {pages} {273} (\bibinfo {year}
  {2020})}\BibitemShut {NoStop}%
\bibitem [{\citenamefont {Honjo}\ \emph {et~al.}(2004)\citenamefont {Honjo},
  \citenamefont {Inoue},\ and\ \citenamefont
  {Takahashi}}]{Honjo2004differential}%
  \BibitemOpen
  \bibfield  {author} {\bibinfo {author} {\bibfnamefont {T.}~\bibnamefont
  {Honjo}}, \bibinfo {author} {\bibfnamefont {K.}~\bibnamefont {Inoue}},\ and\
  \bibinfo {author} {\bibfnamefont {H.}~\bibnamefont {Takahashi}},\ }\bibfield
  {title} {\bibinfo {title} {Differential-phase-shift quantum key distribution
  experiment with a planar light-wave circuit mach--zehnder interferometer},\
  }\href@noop {} {\bibfield  {journal} {\bibinfo  {journal} {Optics letters}\
  }\textbf {\bibinfo {volume} {29}},\ \bibinfo {pages} {2797} (\bibinfo {year}
  {2004})}\BibitemShut {NoStop}%
\bibitem [{\citenamefont {Sibson}\ \emph
  {et~al.}(2017{\natexlab{a}})\citenamefont {Sibson}, \citenamefont {Erven},
  \citenamefont {Godfrey}, \citenamefont {Miki}, \citenamefont {Yamashita},
  \citenamefont {Fujiwara}, \citenamefont {Sasaki}, \citenamefont {Terai},
  \citenamefont {Tanner}, \citenamefont {Natarajan} \emph
  {et~al.}}]{Sibson2017chip}%
  \BibitemOpen
  \bibfield  {author} {\bibinfo {author} {\bibfnamefont {P.}~\bibnamefont
  {Sibson}}, \bibinfo {author} {\bibfnamefont {C.}~\bibnamefont {Erven}},
  \bibinfo {author} {\bibfnamefont {M.}~\bibnamefont {Godfrey}}, \bibinfo
  {author} {\bibfnamefont {S.}~\bibnamefont {Miki}}, \bibinfo {author}
  {\bibfnamefont {T.}~\bibnamefont {Yamashita}}, \bibinfo {author}
  {\bibfnamefont {M.}~\bibnamefont {Fujiwara}}, \bibinfo {author}
  {\bibfnamefont {M.}~\bibnamefont {Sasaki}}, \bibinfo {author} {\bibfnamefont
  {H.}~\bibnamefont {Terai}}, \bibinfo {author} {\bibfnamefont {M.~G.}\
  \bibnamefont {Tanner}}, \bibinfo {author} {\bibfnamefont {C.~M.}\
  \bibnamefont {Natarajan}}, \emph {et~al.},\ }\bibfield  {title} {\bibinfo
  {title} {Chip-based quantum key distribution},\ }\href@noop {} {\bibfield
  {journal} {\bibinfo  {journal} {Nature communications}\ }\textbf {\bibinfo
  {volume} {8}},\ \bibinfo {pages} {13984} (\bibinfo {year}
  {2017}{\natexlab{a}})}\BibitemShut {NoStop}%
\bibitem [{\citenamefont {Ma}\ \emph {et~al.}(2016)\citenamefont {Ma},
  \citenamefont {Sacher}, \citenamefont {Tang}, \citenamefont {Mikkelsen},
  \citenamefont {Yang}, \citenamefont {Xu}, \citenamefont {Thiessen},
  \citenamefont {Lo},\ and\ \citenamefont {Poon}}]{Ma2016silicon}%
  \BibitemOpen
  \bibfield  {author} {\bibinfo {author} {\bibfnamefont {C.}~\bibnamefont
  {Ma}}, \bibinfo {author} {\bibfnamefont {W.~D.}\ \bibnamefont {Sacher}},
  \bibinfo {author} {\bibfnamefont {Z.}~\bibnamefont {Tang}}, \bibinfo {author}
  {\bibfnamefont {J.~C.}\ \bibnamefont {Mikkelsen}}, \bibinfo {author}
  {\bibfnamefont {Y.}~\bibnamefont {Yang}}, \bibinfo {author} {\bibfnamefont
  {F.}~\bibnamefont {Xu}}, \bibinfo {author} {\bibfnamefont {T.}~\bibnamefont
  {Thiessen}}, \bibinfo {author} {\bibfnamefont {H.-K.}\ \bibnamefont {Lo}},\
  and\ \bibinfo {author} {\bibfnamefont {J.~K.}\ \bibnamefont {Poon}},\
  }\bibfield  {title} {\bibinfo {title} {Silicon photonic transmitter for
  polarization-encoded quantum key distribution},\ }\href@noop {} {\bibfield
  {journal} {\bibinfo  {journal} {Optica}\ }\textbf {\bibinfo {volume} {3}},\
  \bibinfo {pages} {1274} (\bibinfo {year} {2016})}\BibitemShut {NoStop}%
\bibitem [{\citenamefont {Sibson}\ \emph
  {et~al.}(2017{\natexlab{b}})\citenamefont {Sibson}, \citenamefont {Kennard},
  \citenamefont {Stanisic}, \citenamefont {Erven}, \citenamefont {O’Brien},\
  and\ \citenamefont {Thompson}}]{Sibson2017integrated}%
  \BibitemOpen
  \bibfield  {author} {\bibinfo {author} {\bibfnamefont {P.}~\bibnamefont
  {Sibson}}, \bibinfo {author} {\bibfnamefont {J.~E.}\ \bibnamefont {Kennard}},
  \bibinfo {author} {\bibfnamefont {S.}~\bibnamefont {Stanisic}}, \bibinfo
  {author} {\bibfnamefont {C.}~\bibnamefont {Erven}}, \bibinfo {author}
  {\bibfnamefont {J.~L.}\ \bibnamefont {O’Brien}},\ and\ \bibinfo {author}
  {\bibfnamefont {M.~G.}\ \bibnamefont {Thompson}},\ }\bibfield  {title}
  {\bibinfo {title} {Integrated silicon photonics for high-speed quantum key
  distribution},\ }\href@noop {} {\bibfield  {journal} {\bibinfo  {journal}
  {Optica}\ }\textbf {\bibinfo {volume} {4}},\ \bibinfo {pages} {172} (\bibinfo
  {year} {2017}{\natexlab{b}})}\BibitemShut {NoStop}%
\bibitem [{\citenamefont {Bunandar}\ \emph {et~al.}(2018)\citenamefont
  {Bunandar}, \citenamefont {Lentine}, \citenamefont {Lee}, \citenamefont
  {Cai}, \citenamefont {Long}, \citenamefont {Boynton}, \citenamefont
  {Martinez}, \citenamefont {DeRose}, \citenamefont {Chen}, \citenamefont
  {Grein} \emph {et~al.}}]{Bunandar2018metropolitan}%
  \BibitemOpen
  \bibfield  {author} {\bibinfo {author} {\bibfnamefont {D.}~\bibnamefont
  {Bunandar}}, \bibinfo {author} {\bibfnamefont {A.}~\bibnamefont {Lentine}},
  \bibinfo {author} {\bibfnamefont {C.}~\bibnamefont {Lee}}, \bibinfo {author}
  {\bibfnamefont {H.}~\bibnamefont {Cai}}, \bibinfo {author} {\bibfnamefont
  {C.~M.}\ \bibnamefont {Long}}, \bibinfo {author} {\bibfnamefont
  {N.}~\bibnamefont {Boynton}}, \bibinfo {author} {\bibfnamefont
  {N.}~\bibnamefont {Martinez}}, \bibinfo {author} {\bibfnamefont
  {C.}~\bibnamefont {DeRose}}, \bibinfo {author} {\bibfnamefont
  {C.}~\bibnamefont {Chen}}, \bibinfo {author} {\bibfnamefont {M.}~\bibnamefont
  {Grein}}, \emph {et~al.},\ }\bibfield  {title} {\bibinfo {title}
  {Metropolitan quantum key distribution with silicon photonics},\ }\href@noop
  {} {\bibfield  {journal} {\bibinfo  {journal} {Physical Review X}\ }\textbf
  {\bibinfo {volume} {8}},\ \bibinfo {pages} {021009} (\bibinfo {year}
  {2018})}\BibitemShut {NoStop}%
\bibitem [{\citenamefont {Zhan}(2009)}]{Zhan2009cylindrical}%
  \BibitemOpen
  \bibfield  {author} {\bibinfo {author} {\bibfnamefont {Q.}~\bibnamefont
  {Zhan}},\ }\bibfield  {title} {\bibinfo {title} {Cylindrical vector beams:
  from mathematical concepts to applications},\ }\href@noop {} {\bibfield
  {journal} {\bibinfo  {journal} {Advances in Optics and Photonics}\ }\textbf
  {\bibinfo {volume} {1}},\ \bibinfo {pages} {1} (\bibinfo {year}
  {2009})}\BibitemShut {NoStop}%
\bibitem [{\citenamefont {Otte}(2020)}]{Otte2020book}%
  \BibitemOpen
  \bibfield  {author} {\bibinfo {author} {\bibfnamefont {E.}~\bibnamefont
  {Otte}},\ }\href@noop {} {\emph {\bibinfo {title} {Structured Singular Light
  Fields}}}\ (\bibinfo  {publisher} {Springer Nature},\ \bibinfo {year}
  {2020})\BibitemShut {NoStop}%
\bibitem [{\citenamefont {Cai}\ \emph {et~al.}(2012)\citenamefont {Cai},
  \citenamefont {Wang}, \citenamefont {Strain}, \citenamefont {Johnson-Morris},
  \citenamefont {Zhu}, \citenamefont {Sorel}, \citenamefont {O’Brien},
  \citenamefont {Thompson},\ and\ \citenamefont {Yu}}]{Cai2012integrated}%
  \BibitemOpen
  \bibfield  {author} {\bibinfo {author} {\bibfnamefont {X.}~\bibnamefont
  {Cai}}, \bibinfo {author} {\bibfnamefont {J.}~\bibnamefont {Wang}}, \bibinfo
  {author} {\bibfnamefont {M.~J.}\ \bibnamefont {Strain}}, \bibinfo {author}
  {\bibfnamefont {B.}~\bibnamefont {Johnson-Morris}}, \bibinfo {author}
  {\bibfnamefont {J.}~\bibnamefont {Zhu}}, \bibinfo {author} {\bibfnamefont
  {M.}~\bibnamefont {Sorel}}, \bibinfo {author} {\bibfnamefont {J.~L.}\
  \bibnamefont {O’Brien}}, \bibinfo {author} {\bibfnamefont {M.~G.}\
  \bibnamefont {Thompson}},\ and\ \bibinfo {author} {\bibfnamefont
  {S.}~\bibnamefont {Yu}},\ }\bibfield  {title} {\bibinfo {title} {Integrated
  compact optical vortex beam emitters},\ }\href@noop {} {\bibfield  {journal}
  {\bibinfo  {journal} {Science}\ }\textbf {\bibinfo {volume} {338}},\ \bibinfo
  {pages} {363} (\bibinfo {year} {2012})}\BibitemShut {NoStop}%
\bibitem [{\citenamefont {Su}\ \emph {et~al.}(2012)\citenamefont {Su},
  \citenamefont {Scott}, \citenamefont {Djordjevic}, \citenamefont {Fontaine},
  \citenamefont {Geisler}, \citenamefont {Cai},\ and\ \citenamefont
  {Yoo}}]{Su2012OAMdevice}%
  \BibitemOpen
  \bibfield  {author} {\bibinfo {author} {\bibfnamefont {T.}~\bibnamefont
  {Su}}, \bibinfo {author} {\bibfnamefont {R.~P.}\ \bibnamefont {Scott}},
  \bibinfo {author} {\bibfnamefont {S.~S.}\ \bibnamefont {Djordjevic}},
  \bibinfo {author} {\bibfnamefont {N.~K.}\ \bibnamefont {Fontaine}}, \bibinfo
  {author} {\bibfnamefont {D.~J.}\ \bibnamefont {Geisler}}, \bibinfo {author}
  {\bibfnamefont {X.}~\bibnamefont {Cai}},\ and\ \bibinfo {author}
  {\bibfnamefont {S.~J.~B.}\ \bibnamefont {Yoo}},\ }\bibfield  {title}
  {\bibinfo {title} {Demonstration of free space coherent optical communication
  using integrated silicon photonic orbital angular momentum devices},\ }\href
  {https://doi.org/10.1364/OE.20.009396} {\bibfield  {journal} {\bibinfo
  {journal} {Opt. Express}\ }\textbf {\bibinfo {volume} {20}},\ \bibinfo
  {pages} {9396} (\bibinfo {year} {2012})}\BibitemShut {NoStop}%
\bibitem [{\citenamefont {Xie}\ \emph {et~al.}(2018)\citenamefont {Xie},
  \citenamefont {Lei}, \citenamefont {Li}, \citenamefont {Qiu}, \citenamefont
  {Zhang}, \citenamefont {Wang}, \citenamefont {Min}, \citenamefont {Du},
  \citenamefont {Li},\ and\ \citenamefont {Yuan}}]{Xie2018ultra}%
  \BibitemOpen
  \bibfield  {author} {\bibinfo {author} {\bibfnamefont {Z.}~\bibnamefont
  {Xie}}, \bibinfo {author} {\bibfnamefont {T.}~\bibnamefont {Lei}}, \bibinfo
  {author} {\bibfnamefont {F.}~\bibnamefont {Li}}, \bibinfo {author}
  {\bibfnamefont {H.}~\bibnamefont {Qiu}}, \bibinfo {author} {\bibfnamefont
  {Z.}~\bibnamefont {Zhang}}, \bibinfo {author} {\bibfnamefont
  {H.}~\bibnamefont {Wang}}, \bibinfo {author} {\bibfnamefont {C.}~\bibnamefont
  {Min}}, \bibinfo {author} {\bibfnamefont {L.}~\bibnamefont {Du}}, \bibinfo
  {author} {\bibfnamefont {Z.}~\bibnamefont {Li}},\ and\ \bibinfo {author}
  {\bibfnamefont {X.}~\bibnamefont {Yuan}},\ }\bibfield  {title} {\bibinfo
  {title} {Ultra-broadband on-chip twisted light emitter for optical
  communications},\ }\href@noop {} {\bibfield  {journal} {\bibinfo  {journal}
  {Light: Science \& Applications}\ }\textbf {\bibinfo {volume} {7}},\ \bibinfo
  {pages} {18001} (\bibinfo {year} {2018})}\BibitemShut {NoStop}%
\bibitem [{\citenamefont {Zhou}\ \emph {et~al.}(2019)\citenamefont {Zhou},
  \citenamefont {Zheng}, \citenamefont {Cao}, \citenamefont {Zhao},
  \citenamefont {Gao}, \citenamefont {Zhu}, \citenamefont {He}, \citenamefont
  {Cai},\ and\ \citenamefont {Wang}}]{Zhou2019ultra}%
  \BibitemOpen
  \bibfield  {author} {\bibinfo {author} {\bibfnamefont {N.}~\bibnamefont
  {Zhou}}, \bibinfo {author} {\bibfnamefont {S.}~\bibnamefont {Zheng}},
  \bibinfo {author} {\bibfnamefont {X.}~\bibnamefont {Cao}}, \bibinfo {author}
  {\bibfnamefont {Y.}~\bibnamefont {Zhao}}, \bibinfo {author} {\bibfnamefont
  {S.}~\bibnamefont {Gao}}, \bibinfo {author} {\bibfnamefont {Y.}~\bibnamefont
  {Zhu}}, \bibinfo {author} {\bibfnamefont {M.}~\bibnamefont {He}}, \bibinfo
  {author} {\bibfnamefont {X.}~\bibnamefont {Cai}},\ and\ \bibinfo {author}
  {\bibfnamefont {J.}~\bibnamefont {Wang}},\ }\bibfield  {title} {\bibinfo
  {title} {Ultra-compact broadband polarization diversity orbital angular
  momentum generator with 3.6$\times$ 3.6 $\mu$m2 footprint},\ }\href@noop {}
  {\bibfield  {journal} {\bibinfo  {journal} {Science advances}\ }\textbf
  {\bibinfo {volume} {5}},\ \bibinfo {pages} {eaau9593} (\bibinfo {year}
  {2019})}\BibitemShut {NoStop}%
\bibitem [{\citenamefont {Chen}\ \emph {et~al.}(2020)\citenamefont {Chen},
  \citenamefont {Xia}, \citenamefont {Shen}, \citenamefont {Gao}, \citenamefont
  {Yan}, \citenamefont {Jiao}, \citenamefont {Dou}, \citenamefont {Tang},
  \citenamefont {Lu},\ and\ \citenamefont {Jin}}]{Chen2020vector}%
  \BibitemOpen
  \bibfield  {author} {\bibinfo {author} {\bibfnamefont {Y.}~\bibnamefont
  {Chen}}, \bibinfo {author} {\bibfnamefont {K.-Y.}\ \bibnamefont {Xia}},
  \bibinfo {author} {\bibfnamefont {W.-G.}\ \bibnamefont {Shen}}, \bibinfo
  {author} {\bibfnamefont {J.}~\bibnamefont {Gao}}, \bibinfo {author}
  {\bibfnamefont {Z.-Q.}\ \bibnamefont {Yan}}, \bibinfo {author} {\bibfnamefont
  {Z.-Q.}\ \bibnamefont {Jiao}}, \bibinfo {author} {\bibfnamefont {J.-P.}\
  \bibnamefont {Dou}}, \bibinfo {author} {\bibfnamefont {H.}~\bibnamefont
  {Tang}}, \bibinfo {author} {\bibfnamefont {Y.-Q.}\ \bibnamefont {Lu}},\ and\
  \bibinfo {author} {\bibfnamefont {X.-M.}\ \bibnamefont {Jin}},\ }\bibfield
  {title} {\bibinfo {title} {Vector vortex beam emitter embedded in a photonic
  chip},\ }\href@noop {} {\bibfield  {journal} {\bibinfo  {journal} {Physical
  Review Letters}\ }\textbf {\bibinfo {volume} {124}},\ \bibinfo {pages}
  {153601} (\bibinfo {year} {2020})}\BibitemShut {NoStop}%
\bibitem [{\citenamefont {Huang}\ \emph {et~al.}(2022)\citenamefont {Huang},
  \citenamefont {Overvig}, \citenamefont {Xu}, \citenamefont {Malek},
  \citenamefont {Tsai}, \citenamefont {Al{\`u}},\ and\ \citenamefont
  {Yu}}]{Huang2022leaky}%
  \BibitemOpen
  \bibfield  {author} {\bibinfo {author} {\bibfnamefont {H.}~\bibnamefont
  {Huang}}, \bibinfo {author} {\bibfnamefont {A.~C.}\ \bibnamefont {Overvig}},
  \bibinfo {author} {\bibfnamefont {Y.}~\bibnamefont {Xu}}, \bibinfo {author}
  {\bibfnamefont {S.~C.}\ \bibnamefont {Malek}}, \bibinfo {author}
  {\bibfnamefont {C.-C.}\ \bibnamefont {Tsai}}, \bibinfo {author}
  {\bibfnamefont {A.}~\bibnamefont {Al{\`u}}},\ and\ \bibinfo {author}
  {\bibfnamefont {N.}~\bibnamefont {Yu}},\ }\bibfield  {title} {\bibinfo
  {title} {Leaky-wave metasurfaces for integrated photonics},\ }\href@noop {}
  {\bibfield  {journal} {\bibinfo  {journal} {arXiv preprint arXiv:2207.08936}\
  } (\bibinfo {year} {2022})}\BibitemShut {NoStop}%
\bibitem [{\citenamefont {Zheng}\ \emph {et~al.}(2023)\citenamefont {Zheng},
  \citenamefont {Zhao},\ and\ \citenamefont {Zhang}}]{Zheng2023versatile}%
  \BibitemOpen
  \bibfield  {author} {\bibinfo {author} {\bibfnamefont {S.}~\bibnamefont
  {Zheng}}, \bibinfo {author} {\bibfnamefont {Z.}~\bibnamefont {Zhao}},\ and\
  \bibinfo {author} {\bibfnamefont {W.}~\bibnamefont {Zhang}},\ }\bibfield
  {title} {\bibinfo {title} {Versatile generation and manipulation of
  phase-structured light beams using on-chip subwavelength holographic surface
  gratings},\ }\href {https://doi.org/doi:10.1515/nanoph-2022-0513} {\bibfield
  {journal} {\bibinfo  {journal} {Nanophotonics}\ }\textbf {\bibinfo {volume}
  {12}},\ \bibinfo {pages} {55} (\bibinfo {year} {2023})}\BibitemShut {NoStop}%
\bibitem [{\citenamefont {White}\ \emph {et~al.}()\citenamefont {White},
  \citenamefont {Su}, \citenamefont {Shahar}, \citenamefont {Yang},
  \citenamefont {Ahn}, \citenamefont {Skarda}, \citenamefont {Ramachandran},\
  and\ \citenamefont {Vučković}}]{White2022OAMEmitter}%
  \BibitemOpen
  \bibfield  {author} {\bibinfo {author} {\bibfnamefont {A.~D.}\ \bibnamefont
  {White}}, \bibinfo {author} {\bibfnamefont {L.}~\bibnamefont {Su}}, \bibinfo
  {author} {\bibfnamefont {D.~I.}\ \bibnamefont {Shahar}}, \bibinfo {author}
  {\bibfnamefont {K.~Y.}\ \bibnamefont {Yang}}, \bibinfo {author}
  {\bibfnamefont {G.~H.}\ \bibnamefont {Ahn}}, \bibinfo {author} {\bibfnamefont
  {J.~L.}\ \bibnamefont {Skarda}}, \bibinfo {author} {\bibfnamefont
  {S.}~\bibnamefont {Ramachandran}},\ and\ \bibinfo {author} {\bibfnamefont
  {J.}~\bibnamefont {Vučković}},\ }\bibfield  {title} {\bibinfo {title}
  {Inverse design of optical vortex beam emitters},\ }\href@noop {} {\bibinfo
  {journal} {Accepted for publication in: ACS Photonics}\ }\BibitemShut
  {NoStop}%
\bibitem [{\citenamefont {Holleczek}\ \emph {et~al.}(2011)\citenamefont
  {Holleczek}, \citenamefont {Aiello}, \citenamefont {Gabriel}, \citenamefont
  {Marquardt},\ and\ \citenamefont {Leuchs}}]{Holleczek2011}%
  \BibitemOpen
\bibfield  {journal} {  }\bibfield  {author} {\bibinfo {author} {\bibfnamefont
  {A.}~\bibnamefont {Holleczek}}, \bibinfo {author} {\bibfnamefont
  {A.}~\bibnamefont {Aiello}}, \bibinfo {author} {\bibfnamefont
  {C.}~\bibnamefont {Gabriel}}, \bibinfo {author} {\bibfnamefont
  {C.}~\bibnamefont {Marquardt}},\ and\ \bibinfo {author} {\bibfnamefont
  {G.}~\bibnamefont {Leuchs}},\ }\bibfield  {title} {\bibinfo {title}
  {Classical and quantum properties of cylindrically polarized states of
  light},\ }\href {https://doi.org/10.1364/OE.19.009714} {\bibfield  {journal}
  {\bibinfo  {journal} {Opt. Express}\ }\textbf {\bibinfo {volume} {19}},\
  \bibinfo {pages} {9714} (\bibinfo {year} {2011})}\BibitemShut {NoStop}%
\bibitem [{\citenamefont {Klyshko}(1988)}]{klyshko1988simple}%
  \BibitemOpen
  \bibfield  {author} {\bibinfo {author} {\bibfnamefont {D.}~\bibnamefont
  {Klyshko}},\ }\bibfield  {title} {\bibinfo {title} {A simple method of
  preparing pure states of an optical field, of implementing the
  einstein--podolsky--rosen experiment, and of demonstrating the
  complementarity principle},\ }\href@noop {} {\bibfield  {journal} {\bibinfo
  {journal} {Soviet Physics Uspekhi}\ }\textbf {\bibinfo {volume} {31}},\
  \bibinfo {pages} {74} (\bibinfo {year} {1988})}\BibitemShut {NoStop}%
\bibitem [{\citenamefont {Siegman}(1986)}]{Siegman1986}%
  \BibitemOpen
  \bibfield  {author} {\bibinfo {author} {\bibfnamefont {A.}~\bibnamefont
  {Siegman}},\ }\href@noop {} {\emph {\bibinfo {title} {Lasers}}}\ (\bibinfo
  {publisher} {University Science Books},\ \bibinfo {year} {1986})\BibitemShut
  {NoStop}%
\bibitem [{\citenamefont {Boyd}\ and\ \citenamefont {Gordon}(1961)}]{Boyd1961}%
  \BibitemOpen
  \bibfield  {author} {\bibinfo {author} {\bibfnamefont {G.~D.}\ \bibnamefont
  {Boyd}}\ and\ \bibinfo {author} {\bibfnamefont {J.~P.}\ \bibnamefont
  {Gordon}},\ }\bibfield  {title} {\bibinfo {title} {Confocal multimode
  resonator for millimeter through optical wavelength masers},\ }\href@noop {}
  {\bibfield  {journal} {\bibinfo  {journal} {Bell System Technical Journal}\
  }\textbf {\bibinfo {volume} {40}},\ \bibinfo {pages} {489} (\bibinfo {year}
  {1961})}\BibitemShut {NoStop}%
\bibitem [{\citenamefont {Goos}(2021)}]{Goos2021}%
  \BibitemOpen
  \bibfield  {author} {\bibinfo {author} {\bibnamefont {Goos}},\ }\href@noop {}
  {\bibinfo {title} {Github repository (accessed 2022-09-09)}},\ \bibinfo
  {howpublished} {\url{https://github.com/stanfordnqp/spins-b}} (\bibinfo
  {year} {2021})\BibitemShut {NoStop}%
\bibitem [{\citenamefont {Su}\ \emph {et~al.}(2020)\citenamefont {Su},
  \citenamefont {Vercruysse}, \citenamefont {Skarda}, \citenamefont {Sapra},
  \citenamefont {Petykiewicz},\ and\ \citenamefont
  {Vu{\v{c}}kovi{\'c}}}]{Su2020nanophotonic}%
  \BibitemOpen
  \bibfield  {author} {\bibinfo {author} {\bibfnamefont {L.}~\bibnamefont
  {Su}}, \bibinfo {author} {\bibfnamefont {D.}~\bibnamefont {Vercruysse}},
  \bibinfo {author} {\bibfnamefont {J.}~\bibnamefont {Skarda}}, \bibinfo
  {author} {\bibfnamefont {N.~V.}\ \bibnamefont {Sapra}}, \bibinfo {author}
  {\bibfnamefont {J.~A.}\ \bibnamefont {Petykiewicz}},\ and\ \bibinfo {author}
  {\bibfnamefont {J.}~\bibnamefont {Vu{\v{c}}kovi{\'c}}},\ }\bibfield  {title}
  {\bibinfo {title} {Nanophotonic inverse design with spins: Software
  architecture and practical considerations},\ }\href@noop {} {\bibfield
  {journal} {\bibinfo  {journal} {Applied Physics Reviews}\ }\textbf {\bibinfo
  {volume} {7}},\ \bibinfo {pages} {011407} (\bibinfo {year}
  {2020})}\BibitemShut {NoStop}%
\bibitem [{\citenamefont {Sapra}\ \emph {et~al.}(2019)\citenamefont {Sapra},
  \citenamefont {Vercruysse}, \citenamefont {Su}, \citenamefont {Yang},
  \citenamefont {Skarda}, \citenamefont {Piggott},\ and\ \citenamefont
  {Vu{\v{c}}kovi{\'c}}}]{Sapra2019inverse}%
  \BibitemOpen
  \bibfield  {author} {\bibinfo {author} {\bibfnamefont {N.~V.}\ \bibnamefont
  {Sapra}}, \bibinfo {author} {\bibfnamefont {D.}~\bibnamefont {Vercruysse}},
  \bibinfo {author} {\bibfnamefont {L.}~\bibnamefont {Su}}, \bibinfo {author}
  {\bibfnamefont {K.~Y.}\ \bibnamefont {Yang}}, \bibinfo {author}
  {\bibfnamefont {J.}~\bibnamefont {Skarda}}, \bibinfo {author} {\bibfnamefont
  {A.~Y.}\ \bibnamefont {Piggott}},\ and\ \bibinfo {author} {\bibfnamefont
  {J.}~\bibnamefont {Vu{\v{c}}kovi{\'c}}},\ }\bibfield  {title} {\bibinfo
  {title} {Inverse design and demonstration of broadband grating couplers},\
  }\href@noop {} {\bibfield  {journal} {\bibinfo  {journal} {IEEE Journal of
  Selected Topics in Quantum Electronics}\ }\textbf {\bibinfo {volume} {25}},\
  \bibinfo {pages} {1} (\bibinfo {year} {2019})}\BibitemShut {NoStop}%
\bibitem [{\citenamefont {Sheridan}\ and\ \citenamefont
  {Scarani}(2010)}]{Sheridan2010}%
  \BibitemOpen
  \bibfield  {author} {\bibinfo {author} {\bibfnamefont {L.}~\bibnamefont
  {Sheridan}}\ and\ \bibinfo {author} {\bibfnamefont {V.}~\bibnamefont
  {Scarani}},\ }\bibfield  {title} {\bibinfo {title} {{Security proof for
  quantum key distribution using qudit systems}},\ }\href
  {https://doi.org/10.1103/PhysRevA.82.030301} {\bibfield  {journal} {\bibinfo
  {journal} {Physical Review A}\ }\textbf {\bibinfo {volume} {82}},\ \bibinfo
  {pages} {030301} (\bibinfo {year} {2010})},\ \Eprint
  {https://arxiv.org/abs/1003.5464} {1003.5464} \BibitemShut {NoStop}%
\bibitem [{\citenamefont {Marrucci}\ \emph {et~al.}(2006)\citenamefont
  {Marrucci}, \citenamefont {Manzo},\ and\ \citenamefont
  {Paparo}}]{Marrucci2006optical}%
  \BibitemOpen
  \bibfield  {author} {\bibinfo {author} {\bibfnamefont {L.}~\bibnamefont
  {Marrucci}}, \bibinfo {author} {\bibfnamefont {C.}~\bibnamefont {Manzo}},\
  and\ \bibinfo {author} {\bibfnamefont {D.}~\bibnamefont {Paparo}},\
  }\bibfield  {title} {\bibinfo {title} {Optical spin-to-orbital angular
  momentum conversion in inhomogeneous anisotropic media},\ }\href@noop {}
  {\bibfield  {journal} {\bibinfo  {journal} {Physical Review Letters}\
  }\textbf {\bibinfo {volume} {96}},\ \bibinfo {pages} {163905} (\bibinfo
  {year} {2006})}\BibitemShut {NoStop}%
\end{thebibliography}
\end{document}